\documentstyle[preprint,aps,eqsecnum]{revtex}
\tightenlines
\begin{document}
\draft
\title{Segment Motion in the Reptation Model \\
       of Polymer Dynamics. I. Analytical Investigation}
\author{U.\ Ebert$^1$, L.\ Sch\"afer$^2$, and A.\ Baumg\"artner$^3$} \address{$^1$ Instituut-Lorentz, Universiteit Leiden, Postbus 9506,         2300 RA Leiden, the Netherlands,}
\address{$^2$ Fachbereich Physik, Universit\"at Essen,
         45117 Essen, Germany,}
\address{$^3$ Institut f\"ur Festk\"orperforschung and Forum
        Modellierung, Forschungszentrum J\"ulich, 52425 J\"ulich, Germany.}
\date{submitted to J.\ Stat.\ Phys.\ on September 18, 1997}
\maketitle
\begin{abstract}

We analyze the motion of individual beads of a polymer chain using a discrete version of De Gennes' reptation model that describes the motion of a
polymer through an ordered lattice of obstacles. The motion within the tube
can be evaluated rigorously, tube renewal is taken into account in an approximation motivated by random walk theory. We find microstructure effects to be present for remarkably large times and long chains,
affecting essentially all present day computer experiments.
The various asymptotic power laws, commonly considered as typical
for reptation, hold only for extremely long chains. Furthermore, for an arbitrary segment even in a very long chain, we find a rich variety of
fairly broad crossovers, which for practicably accessible chain lengths overlap and smear out the asymptotic power laws. Our analysis suggests observables specifically adapted to distinguish reptation from motions dominated by disorder of the environment.

key words: reptation, polymer dynamics
\end{abstract}

\section{Introduction}

The reptation concept, introduced long ago by De Gennes \cite{Z1}, has become a basic scenario of polymer dynamics. It deals with the motion of a
flexible chain surrounded by a network of impenetrable obstacles. Originally
it has been proposed for understanding the diffusion of a polymer through a
gel, but it widely is used also to analyze the dynamics of polymers in a melt
or in a dense solution. A recent review may be found in \cite{Z2}.

The basic idea is best explained for polymer diffusion through a gel that is swollen by some solvent. The gel consists of polymer strands that are chemically linked together to form a three dimensional
network. In the solvent filled space between the strands
some free chain not chemically linked to the gel may exert a Brownian
motion. The hard core interaction with the strands of the gel constrains this motion to a `tube' that is defined by the strands surrounding the instantaneous chain configuration.
Due to crosslinking, these `topological constraints' are essentially fixed. (The thermal motion of the gel is neglected.)
The array of topological constraints forming
a tube around the instantaneous chain configuration is one important ingredient of reptation. The second important ingredient of De Gennes' theory is the specific mechanism of microscopic chain motion within the tube. Since the chain is highly flexible, it cannot move as a whole like a rigid body, but it moves through thermal displacements of
individual segments. It also does not lie stretched in its tube, but as a consequence of flexibility and entropy, there will be a finite density of small wiggles or side loops of `spared length'.
These wiggles, called `defects' by De Gennes, exert a diffusive motion along the chain and thus along the tube. This is taken as the basic dynamical mechanism.

If a defect reaches a chain end, it can be destroyed, thus prolonging the tube by its spared length in some random direction. A chain end can also contract into the tube, thus creating a defect and destroying
the end of the previous tube over a distance of its spared length.
This mechanism results in a thermal motion of the chain ends,
which gradually destroys the original tube until
finally the chain has found a completely new configuration.

Accordingly, the calculation of dynamical correlation functions
in the reptation model can be decomposed into two well separated steps.
The first step is the one-dimensional diffusion of conserved
defects along the chain as the basic thermal motion. This yields
the curvilinear displacement of some interior segment along the tube,
and tube renewal at the chain ends. The second step consists in
embedding the segment displacement into real space by taking
the (equilibrium) random walk configuration of the chain and thus
of the tube into account.

Treating defect dynamics in a continuum approximation, De Gennes \cite{Z1}
showed the existence of several characteristic time scales and derived asymptotic power laws. We here recall some of the results, using notation
adequate for a discrete chain composed of $N$ segments.

There emerge three characteristic time scales. A microscopic time
\begin{eqnarray*}
T_{0} \sim N^{0}
\end{eqnarray*}
is needed for the chain motion to feel the constraining environment. For
shorter time intervals the chain does not properly realize that it is confined to a tube. A second time
\begin{eqnarray*}
T_{2} \sim N^{2}
\end{eqnarray*}
is of the order of the Rouse time, i.e., the longest relaxation time of a
free chain. In the present context it, however, has a quite different interpretation: it is the time a defect needs to diffuse over the whole length of the chain. The third scale
\begin{eqnarray*}
T_{3} \sim N^{3}
\end{eqnarray*}
is known as `reptation time'. Time intervals of the order of $T_{3}$ are
needed to completely destroy the original tube.

We now consider the average displacement in space of some internal bead $j$
of a chain of length $N$:
\begin{equation}
g_{1}(j,N,t) = \left\langle \overline{({\bf r}_{j}(t) - {\bf r}_{j}(0))^{2}}
\right\rangle
\end{equation}
The bar indicates the average over all defect motions along the chain, and
the pointed brackets denote the average over all tube configurations. The
reptation
model predicts
\begin{equation}
g_{1}(j,N,t) \sim\left\{ \begin{array}{cc}
t^{1/4} & T_{0} \ll t \ll T_{2} \\
(t/N)^{1/2} & T_{2} \ll t \ll T_{3} \\
t/N^{2} & T_{3} \ll t \: \: \:. \end{array} \right.
\end{equation}
Note the asymptotic nature of these results. In precise mathematical terms the
first line of Eq.\ (1.2), for instance,
reads (we use the notation $\bar{j} = j/N$):
\begin{eqnarray*}
\lim_{{\scriptsize \begin{array}{cc} t &\rightarrow \infty \\ \bar{j}& \mbox{fixed}
\end{array}} } \left[ t^{-1/4} \lim_{{\scriptsize\begin{array}{cc} N &\rightarrow \infty \\ t,\bar{j}& \mbox{fixed} \end{array}} }\: g_{1} (j,N;t)\right] = const \: \: \:.
\end{eqnarray*}
These predictions therefore concern long chains and times large on microscopic ($T_{0}$) or even mesoscopic ($T_{2},T_{3}$) scales. Similar
results for other quantities like the motion of the center of mass have also
been derived.

Much work has aimed at extracting these power laws from simulation data or
physical experiments on polymer diffusion mostly through melts, dense solutions, gels or spatially fixed, but disordered configurations of obstacles. On a superficial level, the expected behavior is seen, but closer inspection proves the outcome to be inconclusive. For instance, for the central bead of the chain one finds some slowing down of the motion that according to most simulations in some intermediate range seems to follow an effective power law
\begin{eqnarray*}
g_{1} \left(\frac{N}{2},N,t\right) \sim t^{x}, \: \: \: \: \: 0.3 \alt x
\alt 0.4 \: \:.
\end{eqnarray*}
This regime immediately is followed by a smooth crossover towards free diffusion: $g_{1} \sim t$, with no proper indication of a $t^{1/2}$-regime.
For the center of mass motion the results likewise are unsatisfactory.

The interpretation of these results is quite ambiguous. In melts
and dense solutions a slow relaxation of the surrounding medium,
known as `constraint release', will play a role \cite{Z3}. For gels,there exist hints that some types of
disorder may strongly modify the chain dynamics \cite{Z4,Z5,Z6,Z7}. Also a quite trivial effect should be noted: The bead motion of a
simple Rouse chain in a straight tube of finite diameter shows a
crossover from three dimensional motion at short times to one
dimensional motion at larger times, that closely resembles simulational results found in the literature, but that has little to do with reptation.
Nevertheless two results of the simulations deserve special interest here:
Following the change of chain configurations with time, Kremer and Grest
\cite{Z8} convincingly demonstrated that the tube concept is valid even in
melts. Again in a simulation of melts, Shaffer \cite{Z9} did a controlled
study of the influence of chain impenetrability, showing that it strongly
modifies the motion of a test chain.

This situation clearly longs for a detailed study of the pure reptation scenario, suppressing all additional background effects. A first notable
attempt was undertaken by Evans and Edwards \cite{Z10}. Their work, however,
suffered from short chains and poor statistics. In their model they considered the motion of a random walk chain on a cubic lattice
with a second (dual) cubic lattice (with lattice points on the centers of the cubes of the original lattice) representing the obstacles.
The two-dimensional version of the model is shown in Fig.\ 1. We have carried
through new extensive simulations of this model, and our results are described in detail in a second paper \cite{Z11}. Here we only note that we
indeed find a $t^{1/4}$-regime, (as well as indications of a
$t^{1/2}$-regime), but even under the most favourable circumstances it is
fully developed only for quite long chains and surprisingly long times\cite{Z12}. Most of the time regime is covered by broad crossover regions.

To understand these results, it needs a more detailed analytical evaluation
of the reptation model, giving quantitative results for functions like $g_{1}(j,N,t)$ also outside the asymptotic limits. Particular attention should be paid to the time regime $t \alt T_{2}$, since present day simulations for long chains hardly can reach the regime $t \gg T_{2}$. Furthermore we should look for quantities which allow for a clear distinction
between reptation and Rouse-type motion in a disordered environment. The present work is devoted to that task.

In view of the (computer-) experimental situation we also should account for short time effects. Though these effects to some extent will depend on the microscopic model, it for the interpretation of the data is most important to get some feeling for their typical character and range. We thus use a fully discretized version of the original
De Gennes model, adequate for an ordered lattice of obstacles. Depending on the time regime we calculate $g_{1}(j,N,t)$ and related quantities either rigorously or in a seemingly quite good approximation. For large times
some of our results with small modifications also could be extracted from equations given in \cite{Z1}. As shown in the
second paper \cite{Z11} our analytical results perform very well in explaining the Monte Carlo data. They thus provide a firm basis for the study
of polymer motion through more complicated environments like melts
or a disordered background. A short account of some of our analytical and numerical results has been published in \cite{Z12}.

We should delimit our work from two related, but different lines of approach. Rubinstein \cite{Z13} proposed a similar lattice model, that nowadays widely is used in the theory of electrophoresis. Compared to his
model our approach is simpler, since without external fields the defect motion decouples
from the chain configuration. As a result, we can calculate dynamic correlation functions analytically. Doi and Edwards \cite{Z14}, in their work
on melts, retained the tube concept but abstracted from the specific dynamical mechanism of defect motion. They consider the diffusion of a so
called `primitive chain', which is a fictitious object fixing the tube configuration. In the large time regime $t \gg T_{2}$, this approach is equivalent to De Gennes' model, but for shorter times the intrinsic dynamical
mechanism should not be ignored. Doi \cite{Z15} extented this theory by adding internal Rouse dynamics, thus considering a (stretched) one dimensional Rouse chain in a tube. If we ignore the discreteness of the defect motion, our results reduce to those of this model, and for those aspects of the tube renewal specifically considered in \cite{Z15}, we then
find similar results.

This paper is organized as follows. In Sect.\ II we define our model, and we
derive expressions for physical observables. We mainly will consider the
single-bead motion. Quantities like the center of mass, which involve two-bead correlations, in principle are accessible also, but will not be
considered here. In Sect.\ III we derive the generating function of the moments of the single bead motion. It yields rigorous expressions for
quantities like $g_{1}(j,N,t)$, as long as the bead $j$ on average
still sits in a part of the original tube. The analytical and numerical evaluation of our results in that time regime is contained in Sect.\ IV. Sect.\ V deals with end effects, presenting a step towards the description
of tube renewal. The calculation of $g_{1}(0,N,t)$ amounts to determine the average maximal value within the time interval $[0,t]$ of a correlated
stochastic process. For this problem no rigorous solution is known.We use a `mean hopping rate' approximation, that is rigorously valid for uncorrelated, i.e.,  Markovian processes.
In the same spirit, we then evaluate the effect of tube renewal on some arbitrary bead $j$ for times $t < T_{3}$, and we estimate the reptation time
$T_{3}$. Finally Sect.\ VI summarizes our findings. Some useful formulae and
more complicated calculations are presented in appendices.

\section{Model and Observables}
\subsection{Construction}

In constructing our model, we are guided by reptative motion of a lattice
chain. Assume for simplicity that the chain is represented by a random walk
of $N$ steps (segments) on a square lattice of lattice constant $\ell_{0}$.
The configuration is fixed by the bead coordinates $\{{\bf
r}_{0}(t),\ldots,{\bf r}_{N}(t)\}$, which give the positions of the $N+1$
endpoints of all the segments. A lattice of impenetrable obstacles occupies
the centers of the cells of the first lattice, see Fig.\ 1. The obstacles
exclude any kink jumps or more complicated modes of motion, and leave only
tips of hairpins, i.e., beads $j$ with ${\bf r}_{j+1} = {\bf r}_{j-1}$,
free to move. ${\bf r}_{j}$ then can move between all sites neighboring ${\bf r}_{j-1}$. If the hairpin happens to lie down on the chain, it can
stand up on a different bead, and so diffuse along the chain.
If a hairpin (i.e., a `defect') diffuses across a bead, it displaces it by the spared length $2 \ell_{0}$ along the chain configuration. Of course,
also the endpoints are free to move, and at the ends hairpins are created
or destroyed. In our simulations we used the three-dimensional version of
this model, that originally was proposed by Evans and Edwards \cite{Z10}.

Abstracting from this concrete realization, we consider the chain
configuration as a random walk of $N$ segments of fixed length $\ell_{0}$,
that henceforth will be taken as the unit of length. The defects are modelled as particles sitting on the beads of the chain. Each particle represents a defect of spared length $\ell_{s}$. It can hop to each of the two neighbouring beads along the chain with probability $p$. For a given
particle, the other particles are assumed to be just a part of the chain,
so the particles do not interact, and each particle diffuses not along
a primitive chain, but along the whole random walk configuration of
the chain with length $N$, which corresponds to a physical chain of length
$N+\ell_s$. This allows us to take independent averages
over the particle diffusion along the chain and over the random
walk configuration of the chain in embedding space.
We introduce the equilibrium density $\rho_{0}$ of particles
as the average density of particles per bead. The assumptions
of the model will be further discussed in Subsect.\ II.B below.

As pointed out above, at the chain ends (beads $0$ and $N$) defects can be destroyed or created. We can turn this grandcanonical
problem into an easier treatable canonical one by coupling the chain ends to
large virtual particle reservoirs, located at bead $0$ or $N$, respectively.
A particle may hop from bead $1$ $(N-1)$ into reservoir $0$ $(N)$ with probability $p$. It may hop out of the reservoir onto the chain
(from $0$ to $1$, or from $N$ to $N-1$, resp.) with probability $p_{res}$.
We put $M$ particles into the system. Taking $p_{res}\to0$ and $M\to\infty$,
with $p\rho_0=p_{res}M/2$ fixed, each reservoir becomes a source of
particles hopping independently into the chain and also a sink absorbing particles from the chain end, such that the proper particle dynamics at the chain ends is achieved. This canonical formalism allows a very simple calculation of the displacement of some bead
through simple book keeping of the change of particle number on one
side of the bead.

The calculation of the displacement of a bead $j$ thus decomposes
into three steps: $(i)$ Calculate the motion of an individual
particle on the one-dimensional chain. $(ii)$ Calculate the balance
of particles, that have moved over the bead $j$ within the time of
consideration. $(iii)$ Determine from this balance the displacement of the bead in three-dimensional embedding space.

Let us first characterize the motion of the individual particles on the chain.
$P_{1}(j,t)$ is the probability, that a given particle at time $t$ sits on bead $j$, $0 \leq j \leq N$. This probability develops in
(discretized) time according to the equation
\begin{equation}
P_{1}(j,t+1) = \sum_{j' = 0}^{N}\:W_{j j'} P_{1}(j',t) \: \: \:,
\end{equation}
with the hopping matrix for all $0\le j\le N$
\begin{eqnarray}
W_{jj'} &=& (1-2p) \delta_{jj'} + p (\delta_{j,j' + 1} + \delta_{j,j'-1})
~~\mbox{for}~~ 1 \leq j' \leq N-1
\nonumber \\
W_{j 0} &=& p_{res} \delta_{j 1} + (1 - p_{res}) \delta_{j 0}
\nonumber \\
W_{j N} &=& p_{res} \delta_{j N-1} + (1 - p_{res}) \delta_{jN} ~.
\end{eqnarray}
The normalized stationary solution of Eq.\ (2.1) takes the form 
\begin{eqnarray}
P_{1}^{(eq)}(j) &=& (1/p) /\left[(N\!-\!1)/p+2/p_{res}\right]
\nonumber\\
P_{1}^{(eq)}(0) &=& P_{1}^{(eq)}(N) = p P_{1}^{(eq)}(j)/p_{res}
\end{eqnarray}
for $1 \le j \le N-1$. We have defined the equilibrium density of
particles on the chain (beads $1$ to $N-1$) to be $\rho_{0}$, and we thus have for any one of the $M$ particles
\begin{equation}
P_{1}^{(eq)}(j) = \frac{\rho_{0}}{M}, \: \: \: \: \: 1 \leq j \leq N-1~.
\end{equation}
This fixes $p_{res}$
\begin{equation}
p_{res} = \frac{2 \rho_{0}}{M}\:p + O \left(\frac{1}{M^{2}}\right) \: \: \:,
\end{equation}
and results in
\begin{equation}
P_{1}^{(eq)}(0) = P_{1}^{(eq)}(N) = \frac{1}{2} \left(1 -
\frac{N-1}{M}\:\rho_{0} + O \left(\frac{1}{M^{2}}\right)\right)~.
\end{equation}

So far we have defined a standard and readily solved problem of hopping
dynamics. The special features of reptation emerge when we turn to the physical observables, filling in steps $(ii)$ and $(iii)$.
We first consider the motion of a bead $j$ within a time interval short
enough that it stays in the tube defined by the initial configuration $\{r_{i}(0)\}$. (See Fig.\ 2.) At time $t$ the displacement of that bead
{\em along the tube} is given by
\begin{equation}
\ell_{s} n(j,t) = \ell_{s} [n_{-}(j,t) - n_{+}(j,t)] \: \: \:,
\end{equation}
where $n_{\pm}(j,t)$ is the number of particles that have passed over bead $j$ within time interval $[0,t]$ from the `left' ($j' < j$) or
from the `right' ($j' > j$), respectively. (Thus $n(j,0)=0$ by
definition.) Since the tube
configuration itself is a random walk, bead $j$ {\em in real space} is displaced by a random walk of $\ell_{s}|n(j,t)|$ steps. We thus find
\begin{eqnarray}
g_{1}(j,N,t) &=& \left\langle \overline{({\bf r}_{j}(t) - {\bf
r}_{j}(0))^{2}} \right\rangle \nonumber \\
&=& \ell_{s} \:\overline{|n(j,t)|} \: \: \:.
\end{eqnarray}
We here and in the sequel pay no special attention to the situation where at
time $0$ or $t$ a particle sits right on top of bead $j$. In comparison to
experiments such configurations will yield a small time independent
contribution to $g_{1}(j,N,t)$, (which we discuss further in \cite{Z11}).

We now turn to the motion of the ends of the chain. This intrinsically is
the problem of tube renewal. We again consider the situation of Fig.\ 2,
where at time $t$ a part of the original tube still exists that at time
0 was formed by the beads $j_{<} \leq j' \leq j_{>}$. (Note that $j$ is an index of the bead at all times, and of the tube at time $t=0$.)
We concentrate on the motion of chain end $j = 0$. The length $j_{<}$
on which the original tube is destroyed from the left till time $t$,
is determined by the minimal particle occupation of reservoir $j=0$ within time interval $[0,t]$, or equivalently by the maximal contraction of the chain into the tube
\begin{equation}
j_{<} = \ell_{s}\:n_{max}(t)~,
\end{equation}
\begin{equation}
n_{max}(t) = \max_{s \epsilon[0,t]} \: \{n(0,s)\} \geq 0 ~.
\end{equation}
The newly formed piece of the tube at time $t$ has length
$j_{<} - \ell_{s}\:n(0,t)$.
Thus the end bead effectively has been displaced by a random
walk of length $2 j_{<} - \ell_{s}\:n(0,t)$, and we find
\begin{eqnarray}
g_{1}(0,N,t) &=& \left\langle \overline{({\bf r}_{0}(t) - {\bf
r}_{0}(0))^{2}} \right\rangle \nonumber \\
&=& 2 \ell_{s} \:\overline{n_{max}(t)} - \ell_{s}\:\overline{n (0,t)}~.
\end{eqnarray}
Since the particle motion obeys detailed balance, the stochastic process
$n(j,s)$ is invariant under a change of sign, and the last contribution in
Eq.\ (2.11) averages to zero. Our final expression reads
\begin{equation}
g_{1}(0,N,t) = 2 \ell_{s}\:\overline{n_{max}(t)} \: \: \:.
\end{equation}

An end bead is immediately subject to tube renewal. If we now consider some internal bead $j$, $0 < j \leq \frac{N}{2}$, it initially will move within the original tube, but will feel the tube renewal for sufficiently large times. In discussing this superposition of the two modes of motion we assume $t < T_{3}$, such that a part $[j_{<},j_{>}]$ of the original tube still exists. Furthermore we restrict ourselves to $j \ll \frac{N}{2}$, so that bead $j$ is affected by
tube renewal from the closer chain end 0 only.
For a given realization of the stochastic process we have to distinguish two
situations, corresponding to the two cases discussed above:
\begin{itemize}
\item[a)] $j_{<} < \max \{j, j + \ell_{s}\:n(j,t)\}$, \\
i.e., either the initial position or the end position of bead $j$ at time
$t$ is still in the original tube. Then the bead is displaced by a random
walk of $\ell_{s} |n(j,t)|$ steps.
\item[b)] $j_{<} \geq \max\{j, j + \ell_{s}\:n(j,t)\}$. \\
The bead is displaced by a random walk of $(j_{<} - j) + (j_{<} - j - \ell_{s} n (j,t))$ steps.
\end{itemize}
Combining these results and taking averages we find (recall Eq.\ 2.9)
\begin{eqnarray}
g_{1} (j,N,t)
&=&  \ell_{s}\; \overline{|n(j,t)|\;
\Big[ 1 - \Theta \Big(\ell_{s} n_{max}(t) - j\Big)
\Theta \Big(\ell_{s} n_{max}(t) - j - \ell_{s} n (j,t)\Big)\Big]}
\\
& & + \;\ell_{s} \;\overline{
\left(2 n_{max}(t) - n(j,t) - \frac{j}{\ell_{s}}\right)\;
\Theta \Big(\ell_{s} n_{max}(t) - j\Big)\;
\Theta \Big(\ell_{s} n_{max}(t) - j - \ell_{s} n (j,t)\Big) }~,
\nonumber
\end{eqnarray}
where we introduced the discrete $\Theta$-function
\begin{eqnarray}
\Theta(k) = \left\{ \begin{array}{cc}
1 \: \: \: \: \:, & k = 0,1,2,3,\ldots \\
0 \: \: \: \: \:, & k = -1,-2,-3,\ldots \end{array} \right.~.
\end{eqnarray}
It is easily checked that this expression generalizes the special cases discussed above. Eqs.\ (2.8), (2.12), (2.13) are exact expressions in our
reptation model. They hold as long as a part of the initial tube survives,
i.e., for times $t < T_{3}$.

\subsection{Discussion}

Our model is no one-to-one transcription of any model used in simulations.
Some modifications have been mentioned above: The particle moves along an
effective chain of length $N$, the physical chain being longer by the spared
length of the corresponding defect. If a particle sits right on top of bead
$j$, the spatial position of the corresponding bead of the physical chain may
differ from ${\bf r}_{j}$ by a vector of length $\ell_{s}/2$. Also end beads
can move around without forming a defect, a motion not taken into account.
All these effects, however, are relevant for the motion on microscopic time
scales only, and they can easily be corrected for in comparison with experiments. Furthermore our model implicitly assumes a tube of microscopic
diameter $\sim \ell_{0}$, leading to $p T_{0} \approx 1$. We will see, however, that the discrete nature of the motion is felt for much longer times.

The assumption of noninteracting particles may be more serious. Even in a
random walk chain, realistic defects experience a kind of interaction, if
they meet on the chain. They may merge and form a larger defect, which has to decay before the defects can travel on. This corresponds to a
scattering process in the particle picture. In our model the average effect
of these processes is absorbed into the effective hopping rate, which is a
model parameter that sets the elementary time scale. Only anomalously
large fluctuations in the local defect density could invalidate our results.
Since we are concerned with the dynamics under equilibrium conditions, we can
estimate the probability of such fluctuations from the equilibrium configuration of a random walk on a hypercubic lattice. For a simple estimate note that the density of defects is proportional to $1/(2d)$
with $2d$ the coordination number of a hypercubic lattice.
The probability of larger defect structures is found to
decrease exponentially with the number of defects involved. We therefore
expect that the scattering processes do not seriously affect the model.

We also would like to comment on two questions related to the basic philosophy of the approach. Firstly, where in our model does the assumption of a {\em regular} lattice of obstacles come into play? This happens in two ways. If the obstacles are irregularly distributed
in such a way that they occupy some spatial volume otherwise accessible to the polymer, the polymer or parts of it might be confined to
entropic traps. As a consequence the simple random walk type embedding of the tube into real space is invalid. A second effect of disorder is present even for purely topological constraints, which hinder the motion but exclude no finite space for the static polymer configuration:
In such an environment, the random walk configuration of the chain is unchanged, but the mobility of the defects depends on their spatial position. Both effects effectively couple the defect motion to the
chain configuration. 

Secondly, we comment on our notion of the tube. In a lattice model like that described at the beginning of this section and in Fig.\ 1, the tube may be defined as the non-reversal random walk
configuration of the chain constructed by cutting off all hairpins, double hairpins, etc. Since, however, defects have to move along the whole chain, not only along its backbone, the thus constructed tube has no immediate relevance for the microscopic motion. We thus here use the word `tube' as synonymous to the instantaneous full random walk
configuration of the chain. This definition allows us to proceed much further in our analytical approach. 

We finally comment on the generalization of our model to self-repelling
(excluded volume) chains. First of all, if the reptation model is
applied to melts or dense solutions, it is typically argued, that
the self-repulsion of the individual chain is screened by the presence
of the other chains, such that each chain, though repelling, has a
random walk configuration over longer distances. In an regular array of fixed obstacles, however, self-repulsion has nontrivial effects.
Technically, we have to distinguish the effect of the repulsion
among beads spaced a small or a large distance along the chain. On small
distances, $|j_{1}-j_{2}| = O(1)$, the repulsion among beads $j_{1},j_{2}$
will change the precise physical realization of the defects as well as their
density, and it will lead to a readjustment of the effective hopping rate,
similar to the scattering processes addressed above. This consideration
also applies to models for melts etc. The nontrivial aspects
of the excluded volume problem, however, are due to interactions of beads spaced far
along the chain. Now it is known that the excluded volume interaction decreases the probability of close encounter of such pairs of beads, to the
extent that the number of close encounters, appropriately corrected for the
effects of pairs close along the chain, vanishes for an infinitely long chain
\cite{Z16}. The excluded volume therefore should have no nontrivial influence
on the defect dynamics. Only the static configuration of the chain must respect excluded volume statistics. We thus conclude that Eq.\ (2.7), for instance, for an excluded volume chain in a regular obstacle lattice to first approximation should be replaced by
\begin{equation}
g_{1}(j,N,t) = \ell_{s}\:\left(\overline{|n(j,t)|}\right)^{2\nu} \: \: \:,
\end{equation}
where $\nu \approx 0.588$ is the excluded volume (Flory) exponent. A more
accurate treatment should build upon the known internal correlation functions of the excluded volume problem. Our other expressions are subject to similar modifications.

\section{Generating function and moments}

To calculate the thermal displacement $g_{1} (j,N,t)$ in space
of a bead $j$ of the chain, we have to control the stochastic variables $n(j,t)$ and $n(0,s)$ for $1 \leq s \leq t$ of particle motion along the chain. We thus introduce the distribution function
\begin{equation}
\hat{{\cal P}} (\{n_{s}\},n;j,t) =
\overline{\delta_{nn(j,t)}\:\prod^{t}_{s=1}\:\delta_{n_{s} n(0,s)}} \: \: \:,
\end{equation}
as well as the generating function
\begin{eqnarray}
{\cal P} (\{x_{s}\},x;j,t) &=& \sum^{+ \infty}_{n = - \infty}\: \sum^{+ \infty}_{\{n_{s}\} = - \infty}\:x^{n}\:\prod^{t}_{s=1}\:x_{s}^{n_{s}} \hat{{\cal P}} (\{n_{s}\},n;j,t)
\nonumber \\
&=& \overline{x^{n(j,t)}\:\prod^{t}_{s=1}\:x^{n(0,s)}} \: \: \:.
\end{eqnarray}

To find a manageable expression for $n(j,t)$, we now take advantage of the
fact that by construction of the model, the number of particles is conserved.
We thus can calculate $n(j,t)$ as the change in the number of particles found
at positions $j' \leq j$:
\begin{equation}
n(j,t) = \sum^{M}_{m=1} \Big[\Theta\big(j-j(m,0)\big) -
\Theta\big(j-j(m,t)\big)\Big]~.
\end{equation}
This expression is valid for all $j$ including $j = 0$ and $N$.
$j(m,s)$ is the bead where the $m^{th}$ particle sits at time $s$.
Eq.\ (3.3) thus expresses $n(j,t)$ in terms of single-particle coordinates.

Since the $M$ particles move independently, we now can rewrite Eq.\ (3.2) as
\begin{equation}
{\cal P} (\{x_{s}\},x;j,t) = \left[ Q (\{x_{s}\},x;j,t)\right]^{M} \: \: \:,
\end{equation}
where $Q(\ldots)$ is a single particle contribution.
\begin{equation}
Q (\{x_{s}\},x;j,t) = \overline{x^{\Theta(j-j_{0}) - \Theta (j-j_{t})} \prod^{t}_{s=1}\: x_{s}^{\delta_{0j_{0}} - \delta_{0j_{s}}}}
\end{equation}
(Note that from Eq.\ (2.14), $\Theta (0 - j_{s}) = \delta_{0j_{s}},\:j_{s} \geq
0$). Here $j_{s}$ is the position on the chain where the particle
considered sits at time $s$. Writing out the average explicitly,
we have to evaluate the expression
\begin{eqnarray}
Q (\{x_{s}\},x;j,t) &=& \sum^{N}_{\{j_{0},j_{1},\ldots,j_{t}\} =
0}\:x^{\Theta(j-j_{0}) - \Theta (j-j_{t})} \prod^{t}_{s=1}\:
x_{s}^{\delta_{0j_{0}} - \delta_{0j_{s}}}
\nonumber \\
&\cdot& \prod^{t}_{s'=1}\:W_{j_{s'}j_{s'-1}}\: P_{1}^{(eq)} (j_{0})
\end{eqnarray}

Aiming at the limit $M \rightarrow \infty$, we need to calculate $Q(\ldots)$ including terms of order $1/M$. Since both the hopping rate $p_{res}$ from a reservoir onto the chain and the probability
$P_{1}^{(eq)}(j_{0})$ with $1 \leq j_{0} \leq N-1$ to find the particle on the chain initially, are of order $1/M$ (cf.\ Eqs.\ (2.4), (2.5)), we need to consider two types of processes only:
\begin{itemize}
\item[i)] The particle sits on the chain at time $t = 0$. If during its walk
it is caught by a reservoir, it is not allowed to leave it again.
\item[ii)] At $t = 0$ the particle sits in a reservoir. It may leave it, but
if its caught by a reservoir again, it may not jump on the chain a second
time.
\end{itemize}
The contributions of these processes are easily expressed in terms of the
Green's function for the particle walking on the chain.
\begin{equation}
G_{jj'}^{(0)}(s) = \left(\hat{W}^{s}\right)_{jj'}, \: \: \: \: \: 1 \leq j,
j' \leq N-1
\end{equation}
Here $\hat{W}_{jj'}$ is the matrix $W$ (Eq.\ (2.2)), restricted to the subspace $1 \leq j,j' \leq N-1$. To give an example we consider a special
process of type ii): namely that the particle sits in reservoir 0
at time 0, and on some bead $j'$ with $j<j'<N$ at time $t$, i.e.,
still on the chain, but on the other side of bead $j$.
The contribution of all such processes to Eq.\ (3.5) reads
\begin{eqnarray*}
\lefteqn{\sum_{s=1}^{t} \sum^{N-1}_{j' = j + 1}\: x \: \prod^{t}_{s' = s}\:x_{s'}\: P^{(eq)}(0) \left(W_{00}\right)^{s-1}\:p_{res}
G_{j' 1}^{(0)}(t-s) =}
\nonumber \\
& & \qquad\qquad\qquad\qquad\frac{\rho_{0}
p}{M}\:\sum^{t}_{s=1}\:\sum^{N-1}_{j' = j + 1}\:x\:\prod^{t}_{s' = s}\:x_{s'}\:G^{(0)}_{j' 1}(t-s) + O \left(\frac{1}{M^{2}}\right) \: \: \:,
\end{eqnarray*}
where in the last line we used Eqs.\ (2.2), (2.5), (2.6). The factor of $x$
accounts for the fact that $j_{0} = 0 < j$, whereas $j_{t} = j' > j$. Furthermore, after the particle has left the reservoir, each time step $s'$
contributes a factor $x_{s'}$. All other possible processes can be evaluated
in the same way. The calculation is straightforward. The expressions can be
simplified by using the well known explicit form of $G_{jj'}^{(0)}(s)$:
\begin{equation}
G_{jj'}^{(0)}(s) = \frac{2}{N}\:\sum^{N-1}_{k=1}\:\sin\:\left(\frac{\pi k}{N}\:j\right)\:\sin\:\left(\frac{\pi k}{N}\:j' \right) \alpha_{k}^{s}
\end{equation}
\begin{equation}
\alpha_{k} = 1 - 4 p\:\sin^{2} \left(\frac{\pi k}{2N}\right)
\end{equation}
We here only summarize the final results for (3.5):
\begin{equation}
Q (\{x_{s}\},x;j,t) = 1 + \frac{1}{M} \Big[Q_{1} (x;j,t) + Q_{2}(\{x_{s}\};t)
+ Q_{3}(\{x_{s}\},x;j,t)\Big]~,
\end{equation}
\begin{eqnarray}
Q_{1} (x;j,t) &=& \rho_{0} \left(x + \frac{1}{x} - 2\right) A_{1} (j,t)~,
\\
A_{1} (j,t) &=& \frac{pt}{N} + \frac{1}{2N} \sum_{k=1}^{N-1}\:(1 - \alpha_{k}^{t}) \;\frac{\cos^{2}\left(\frac{\pi k}{N} \left(j +
\frac{1}{2}\right)\right)}{\sin^{2} \left(\frac{\pi k}{2N}\right)}~,
\end{eqnarray}
\begin{eqnarray}
Q_{2} (\{x_{s}\};t) &=& \rho_{0} p \sum^{t}_{s=1}
\left(\prod^{t}_{s'=s}\:x_{s'}^{-1} + \prod^{t}_{s'=s}\:x_{s'} - 2\right)
\\
& & +\; \rho_{0} p^{2} \sum_{0<s_{1}\leq s_{2}<t}
\left(\prod^{t}_{s=s_{1}\rule{0mm}{1ex}}\:x_{s} - 1 \right)
\left(\prod^{t}_{s=s_{2}+1}\:x_{s}^{-1} - 1\right) A_{2}(s_{2} - s_{1})~,
\nonumber \\
A_{2}(s) &=& \frac{2}{N} \sum^{N-1}_{k=1}\:\alpha_{k}^{s}\:\sin^{2} \left(\frac{\pi k}{N}\right)~,
\end{eqnarray}
\begin{eqnarray}
Q_{3} (\{x_{s}\},x;j,t) &=& \rho_{0} p \sum^{t-1}_{s=1}\:\left[(x-1) \left(\prod^{t}_{s'=t-s}\:x_{s'} - 1\right)\right.
\nonumber \\
&& +\;\left. (x^{-1} -1) \left(\prod^{t}_{s'=s+1}\:x_{s'}^{-1} - 1\right)
\right] A_{3}(j,s)~,
\\
A_{3}(j,s) &=& \frac{1}{N} + \frac{2}{N} \sum^{N-1}_{k=1}\:
\alpha_{k}^{s}\:\cos\left(\frac{\pi k}{2N}\right) \;
\cos \left(\frac{\pi k}{N} \left(j + \frac{1}{2}\right)\right)~.
\end{eqnarray}
Some useful properties of the coefficients $A_{b}, b = 1,2,3,$ are
collected in appendix A. Substituting Eq.\ (3.10) into Eq.\ (3.4)
and taking $M \rightarrow \infty$ we find
\begin{equation}
{\cal P} (\{x_{s}\},x;j,t) = \exp \Big\{Q_{1} (x;j,t) + Q_{2}(\{x_{s}\};t)
+ Q_{3} (\{x_{s}\},x;j,t)\Big\}
\end{equation}

We have written this result in a form that factorizes into a contribution
$Q_1$ giving the statistics of $n(j,t)$, a contribution $Q_2$ giving the
statistics of the occupation $\{n(0,s)\}$ of reservoir 0, and an
interference term $Q_{3}$. Note that summation over some variable $n_{s}$ or $n$ in $\hat{{\cal P}}(\{n_{s}\},n;j,t)$, simply amounts to setting the corresponding variable $x_{s}$ or $x$ in
${\cal P}(\{x_{s}\},x;j,t)$ to 1.

At this point we can formally establish the connection to the model of a Rouse chain
in a tube \cite{Z15}. We write
$x = e^{i \varphi}, x_{s} = e^{i \varphi_{s}}$ and we expand to second order
in $\varphi,\varphi_{s}$ to find
\begin{eqnarray}
Q_{1}+Q_{2}+Q_{3} \to
\tilde{Q}_{\varphi} (\{\varphi_{s}\},\varphi;j,t)
&=&
\rho_{0}\; \varphi^{2} \;{\cal G}_{jj}(t)
\nonumber \\
&& ~ -\; \frac{\rho_{0}}{2}\:\sum^{t}_{s_{1},s_{2}=1}\:\varphi_{s_{1}} \varphi_{s_{2}} \left({\cal G}_{00}(s_{1}) + {\cal G}_{00}(s_{2}) - {\cal
G}_{00}(|s_{2} - s_{1}|)\right)
\nonumber \\
&&~ -\; \rho_{0}\:\varphi\:\sum^{t}_{s=1}\:\varphi_{s}
\left( {\cal G}_{0j}(s) +  {\cal G}_{0j}(t) - {\cal G}_{0j}(t-s)\right)~,
\end{eqnarray}
where
\begin{equation}
{\cal G}_{jj'}(s) = \frac{p s}{N} + \frac{1}{2 N}\:\sum^{N-1}_{k=1}\:(1 -
\alpha_{k}^{s})\; \frac{\cos \left(\frac{\pi k}{N} \left(j +
\frac{1}{2}\right)\right) \;\cos \left(\frac{\pi k}{N} \left(j' +
\frac{1}{2}\right)\right)}{\sin^{2} \left(\frac{\pi k}{2 N}\right)} \: \:.
\end{equation}
We here used Eqs.\ (A.7), (A.12) of appendix A to evaluate summations over
$A_{2}(s)$ and $A_{3}(j,s)$. In the limit $s \to \infty$ and $ps$ fixed, ${\cal G}_{jj'}(s)$ reduces to the propagator of the bead motion in a one
dimensional Rouse chain of length $N-1$. For the moments which can be derived
from $\tilde{Q}_{\varphi}$, we in this limit recover the results of the model of a Rouse chain in a tube.

Returning to our full expressions we note that the simplest quantity of interest to us in the sequel, is the distribution of $n(j,t)$, irrespective of
the occupation of the reservoirs. Putting $x_{s} = 1$ for all
$1 \leq s \leq t$, we
find
\begin{equation}
\hat{{\cal P}}_{1}(n;j,t) = \frac{1}{2 \pi
i}\:\oint\:\frac{dx}{x}\:x^{-n}\:\exp (Q_{1}(x;j,t)) \: \:.
\end{equation}
$Q_{1}(x;j,t)$ (Eq.\ 3.11) is invariant under the substitution $x
\rightarrow 1/x$, which leads to
\begin{equation}
\hat{{\cal P}}_{1}(n;j,t) = \hat{{\cal P}}_{1} (-n;j,t) \: \: \:.
\end{equation}
This expresses detailed balance to hold in our equilibrium problem. (Of
course, also invariance under the substitution $j \rightarrow N-j$ holds,
which reflects the symmetry of the chain.) With the substitution $x = e^{i
\varphi}$, we may express $\hat{{\cal P}}_{1}(n;j,t)$ in terms of a modified
Bessel function of first kind $I_{n}(z)$:
\begin{equation}
\hat{{\cal P}}_{1}(n;j,t) = e^{- 2 \rho_{0} A_{1}(j,t)}\:I_{n}(2 \rho_{0}
A_{1}(j,t)) \: \: \:.
\end{equation}
For large values of $\rho_{0} A_{1}(j,t)$, we find
$\overline{n^{2}} \sim \rho_{0} A_{1}(j,t)$, and $\hat{{\cal P}}_{1}$ tends to a Gaussian
\begin{equation}
\hat{{\cal P}}_{1}(n;j,t) \rightarrow (4 \pi \rho_{0}
A_{1}(j,t))^{-1/2}\:\exp\:\left(- \frac{n^{2}}{4 \rho_{0} A_{1}(j,t)}\right)\:
\: \:.
\end{equation}
We observe that this limit is equivalent to the quadratic approximation (3.18) discussed above.

With the explicit form (3.11) of $Q_{1}$, we immediately can
calculate all the cumulants of our distribution
\begin{eqnarray}
\overline{(n(j,t))^{2\ell}}^{c} &=&
{\left(x\:\frac{d}{dx}\right)^{2\ell}}_{x=1}\:\ln\:{\cal P}_{1}(x;j,t)
= 2 \rho_{0} A_{1}(j,t)
\nonumber \\ &\equiv& \overline{n^{2}(j,t)}~.
\end{eqnarray}
The cumulants $\overline{(n(j,t))^{2\ell}}^{c}$ are thus independent of $\ell$ for all integer $\ell$. 
Odd moments, of course, vanish. We, however, are interested also in
odd powers of absolut values. Specifically we need to calculate
\begin{eqnarray}
\overline{|n(j,t)|} &=& \sum_{n = - \infty}^{+ \infty}\:|n| \;\hat{{\cal
P}}_{1}(n;j,t)
\nonumber \\
&=& \frac{1}{\pi i}\:\oint_{|x| < 1}\:dx\:\frac{\exp[Q_{1}(x;j,t)]}{(x-1)^{2}}
\end{eqnarray}
With the substitution $x = \frac{y}{2} + 1 - \frac{1}{2}\:\sqrt{y (y+4)}$ it
is a standard exercise to reduce this expression to the form
\begin{equation}
\overline{|n(j,t)|} = \frac{2}{\sqrt{\pi}}\:\left(\rho_{0}
A_{1}(j,t)\right)^{1/2} \left[ 1 - F_{1} (4 \rho_{0} A_{1} (j,t))\right]
\end{equation}
\begin{equation}
F_{1}(z) = \frac{1}{2\sqrt{\pi}}\:\int_{0}^{z}\:dx\:x^{-3/2}\:e^{-x} \left(\left(1 - \frac{x}{z}\right)^{-1/2} -1\right) - \frac{1}{2
\sqrt{\pi}}\:\Gamma \left(- \frac{1}{2},z\right)\: \:.
\end{equation}
Here $\Gamma(\alpha,z)$ denotes the incomplete $\Gamma$-function. Similar
expressions can be derived for higher odd moments $\overline{|n(j,t)|^{2\ell
+1}}$.

Starting with $F_{1}(0) = 1$, the function $F_{1}(z)$ decreases
monotonically, and asymptotically it behaves as
\begin{equation}
F_{1}(z) = \frac{1}{4z} + O \left(\frac{1}{z^{2}}\right) \: \: \:.
\end{equation}
Since for $t \rightarrow \infty$, the function $A_{1}(j,t) \sim
\rho_0 t \rightarrow \infty$, the contribution of $F_1$ in
this limit can be neglected, and we find the asymptotic result
\begin{equation}
\overline{|n(j,t)|}
\approx \frac{2}{\sqrt{\pi}} \left(\rho_{0} A_{1}(j,t)\right)^{1/2}
= \sqrt{\frac{2}{\pi}}\:\:\: \overline{(n^{2}(j,t))}^{1/2} \: \: \:.
\end{equation}
In the next section we will find that $A_{1}(j,t)$ in a large initial region increases quite slowly with $t$, so that the simple relation (3.29) holds only for rather large time. As we will see, this may be one
reason for the failure of previous simulations to exhibit the
$t^{1/4}$-behavior.

For a Gaussian distributed {\em continuous} variable Eq.\ (3.29) is rigorous. The correction function $F_{1}(z)$ essentially reflects the fact
that $n(j,t)$ ranges over a {\em discrete} set of integers. This easily is
checked by calculating $\overline{|n(\Gamma)|}$ for a process, Gaussian distributed with width $\Gamma$, but restricted to the integers. Depending on
$\Gamma$, we find corrections to the result (3.29) which numerically are
quite similar to the effect of $F_{1}(z)$.

For a discussion of the general motion of some segment and its interference
with tube renewal, it is of interest to consider the correlation among $n(j,t)$ and $n(0,t)$. Eq.\ (3.2) yields
\begin{eqnarray}
\hat{{\cal P}}_{2}(n_{t},n;0,j,t) &=&
\frac{1}{(2\pi i)^{2}}\:\oint\:\frac{dx_{t}}{x_{t}}\:\frac{dx}{x}
\:x_{t}^{-n_{t}}\:x^{-n}\:{\cal P}(\{1,\ldots1,x_{t}\},x;j,t)
\\
&=& \frac{1}{4\pi^{2}}\:\int_{-\pi}^{\pi}\:d \varphi_{t}\:d \varphi\:e^{-
i n_{t} \varphi_{t} - i n \varphi}
\cdot \exp \Bigg\{2 \rho_{0}(\cos\:\varphi_{t} -1)\left[ A_{1}(0,t) - \tilde{A}_{3}(j,t)\right]
\nonumber \\
& & ~~+ 2 \rho_{0}(\cos\:\varphi -1) \left[ A_{1}(j,t) -
\tilde{A}_{3}(j,t)\right]  + 2 \rho_{0}(\cos\:(\varphi_{t} + \varphi) -1)
\tilde{A}_{3}(j,t)\Bigg\}~.
\nonumber
\end{eqnarray}
We here introduced
\begin{equation}
\tilde{A}_{3}(j,t) = p \left(\frac{t-1}{N} +
\sum_{s=1}^{t-1}\:A_{3}(j,s)\right) \: \: \:,
\end{equation}
and we used Eq.\ (A.7) to evaluate $Q_{2}(\{1,\ldots,1,x_{t}\},t)$. The distribution $\hat{{\cal P}}_{2}$ changes its character for $t \approx j^{2}$. At short times $t \ll j^{2}, \tilde{A}_{3}(j,t)$ vanishes like $e^{-j^{2}/4 p t}$ (cf.\ Eq.\ A.16), and
$\hat{{\cal P}}_{2}$ factorizes.
\begin{equation}
\hat{{\cal P}}_{2}(n_{t},n;0,j,t) = \hat{{\cal P}}_{1}(n_{t},0,t)\;\hat{{\cal P}}_{1}(n,j,t)~~~\mbox{ for }~ t/j^{2} \to 0 ~.
\end{equation}
As expected the two beads 0 and $j$ move independently. In the
opposite limit of $t \gg j^{2}$, $\tilde{A}_{3}(j,t)$ diverges proportional
to $t$ (cf.\ Eq.\ A.12), whereas the combinations
$A_{1}(j',t) - \tilde{A}_{3}(j,t)$ with $j' = 0$ or $j$, saturate at some
$j$-dependent value. $\hat{{\cal P}}_{2}$ factorizes into a Gaussian distribution of $n_{+} = (n_{t}+n)/2$ and a distribution of
$n_{-} = n_{t}-n$:
\begin{equation}
\hat{{\cal P}}_{2}(n_{t},n;0,j,t) = \left(4 \pi \rho_{0}
\tilde{A}_{3}(j,t)\right)^{- 1/2}\:\exp \left(- \frac{n_{+}^{2}}{4 \rho_{0}
\tilde{A}_{3}(j,t)}\right) \hat{{\cal P}}_{1}^{(-)}(n_{-},j,t)
~~~\mbox{ for }~  t \gg j^{2}~.
\end{equation}
$\hat{{\cal P}}_{1}^{(-)}$ takes the form characteristic for a single bead
process (cf.\ Eq.\ 3.22).
\begin{equation}
\hat{{\cal P}}_{1}^{(-)}(n_{-},j,t) = e^{- \Delta(j,t)} I_{|n_{-}|}(\Delta(j,t))
\end{equation}
\begin{eqnarray}
\Delta(j,t) &=& j - \frac{2}{N} \sum_{k=1}^{N-1}\:\alpha_{k}^{t}\:\left[
\frac{\sin \left(\frac{\pi k}{2N} (j+1)\right) \;\sin \left(\frac{\pi k}{2N}\:j\right)}{\sin \left(\frac{\pi k}{2N}\right)}\right]^{2}
\nonumber \\
&\to& j ~~~\mbox{ for }~  t \gg j^{2}~.
\end{eqnarray}
The result (3.33)-(3.35) expresses the fact that most of the defects created
by chain end $0$ and not destroyed again at this end by time $t$, for $t \gg
j^{2}$ have diffused over bead $j$ without returning, so that the stochastic variable $n(j,t)$ then
closely follows $n(0,t)$. For $t \gg N^{2}$ it in particular implies that
the total chain motion is dominated by the single stochastic process $n(0,t)$. This observation underlies all theories of the large time limit
\cite{Z1,Z14}.

\section{Motion inside the tube}

Within our model, we now can treat the thermal displacement of some bead $j$ exactly as long as $t$ is so small that the tube renewal has not reached that bead. Eqs.\ (2.8) and (3.26) yield
\begin{eqnarray}
g_{1}(j,N,t) &=& \left\langle \overline{({\bf r}_{j}(t) - {\bf r}_{j}(0))^{2}}
\right\rangle
\nonumber \\
&=& \ell_{s} \sqrt{\frac{2}{\pi}\:\overline{n^{2}(j,t)}} \left(1 - F_{1}(2
\overline{n^{2}(j,t)})\right)
\end{eqnarray}
\begin{equation}
\overline{n^{2}(j,t)} = 2 \rho_{0} A_{1}(j,t)
\end{equation}
In view of the symmetry of the chain we henceforth restrict the analysis to $0
\leq j \leq \frac{N}{2}$. We now evaluate several limits of $g_{1}(j,N,t)$ analytically.
\begin{itemize}
\item[A)] Introducing the variables
\begin{equation}
\bar{t} = \frac{p t}{N^{2}}, \: \: \: \:
\bar{j} = j/N \: \: \: \:,
\end{equation}
we take the limit of long chains $N \rightarrow \infty$, and we consider intermediate times $0 < \bar{t} < \infty$, i.e., times of order $T_{2}$. We thus focus on the regime $T_{0} \ll t \ll T_{3}$, where the crossover from $g_{1}
\sim t^{1/4}$ to $g_{1} \sim t^{1/2}$ occurs. (See Eq.\ (1.2).) It will be shown in Sec.~V.C, that for $\bar{j} > 0$ it needs a time $\bar{t} = \bar{T}_{R}(\bar{j})  \rightarrow \infty$, before the tube is destroyed up
to bead $j = \bar{j} N \rightarrow \infty$. Thus  in the present
limit tube renewal is irrelevant and
our results are exact. 

It now is useful to rewrite $A_{1}(j,t)$ as (cf.\ Appendix A.1)
\begin{equation}
A_{1}(j,t) = N \bar{A}_{1}(\bar{j},\bar{t})
\end{equation}
\begin{eqnarray}
\bar{A}_{1}(\bar{j},\bar{t}) &=& \bar{t} + \left(\bar{j} -
\frac{1}{2}\right)^{2} + \frac{1}{12}
\\
& & - \frac{1}{2N^{2}} \sum^{N-1}_{k=1}\:\exp
\left[\frac{N^{2}}{p}\:\bar{t}\:\ln\:\left(1 - 4 p \:\sin^{2} \left(\frac{\pi
k}{2N}\right)\right)\right] \frac{\cos^{2}\left(\pi k \bar{j} + \frac{\pi
k}{2N}\right)}{\sin^{2} \left(\frac{\pi k}{2N}\right)} \nonumber 
\end{eqnarray}
Taking $N \rightarrow \infty$, we find the limiting form of
$\bar{A}_{1}(\bar{j},\bar{t})$:
\begin{equation}
\bar{A}_{1}(\bar{j},\bar{t}) = \bar{t} + \left(\bar{j} -
\frac{1}{2}\right)^{2} + \frac{1}{12} -
\frac{2}{\pi^{2}}\:\sum^{\infty}_{k=1}\:\frac{\cos^{2}(\pi k
\bar{j})}{k^{2}}\:e^{- \pi^{2} k^{2} \bar{t}}
\end{equation}
\begin{equation}
= \frac{1}{2}
\sqrt{\frac{\bar{t}}{\pi}}\:\int_{0}^{1}\:\frac{ds}{\sqrt{s}}\:\sum^{+ \infty}_{\nu = - \infty}\:\left[e^{- \frac{\nu^{2}}{\rule{0mm}{1ex}\bar{t}s}}
+ e^{- \frac{(\nu - \bar{j})^{2}}{\rule{0mm}{1ex}\bar{t} s}}\right]
\end{equation}
Eq.\ (4.7) results from a Poisson transformation and is easily expressed in
terms of error functions.

Clearly $A_{1}(j,t)$ diverges for $N \rightarrow \infty,\:\bar{j},\bar{t}$ fixed.
Therefore the correction function $F_{1}(\ldots)$ in (4.1) vanishes, and we find
\begin{eqnarray}
\bar{g}_{1}(\bar{j},\bar{t}) &=& \lim_{{\scriptsize\begin{array}{cc} N &\rightarrow \infty \\[-0.4ex] \bar{t},\bar{j}& \mbox{fixed} \end{array}} }\:
N^{- 1/2} g_{1}(j,N,t)
\nonumber \\
&=& 2 \ell_{s} \left(
\frac{\rho_{0}}{\pi}\:\bar{A}_{1}(\bar{j},\bar{t})\right)^{1/2}
\end{eqnarray}

For chains of length $N$ finite, but so large that they can be described by these limiting expressions, we obviously can absorb all microscopic model
parameters into a rescaling of $N$ and $t$
\begin{eqnarray}
N \rightarrow \tilde{N} &=& \ell_{s}^{2}\:\rho_{0}\:N
\nonumber \\
t \rightarrow \tilde{t} &=& \ell_{s}^{4}\:\rho_{0}^{2}\:p t
\end{eqnarray}
Thus the present results are independent of the microstructure and can be
written in universal form as
\begin{equation}
g_{1}(j,N,t) = \tilde{t}^{1/4} \tilde{g}_{1}
\left(\frac{j}{N},\frac{\tilde{t}}{\tilde{N}^{2}}\right) \: \: \:.
\end{equation}
They also can be derived from a continuous model and agree with the result
of ref. \cite{Z1}, appendix B, except that in that work $g_{1}$ was identified with $\left(\overline{n^{2}(j,t)}\right)^{1/2}$, in our language.
This introduces an additional factor $\sqrt{\pi/2}$.

Considering now extreme values of $\bar{t}$ we from Eqs.\ (4.6), (4.7) find
the power laws
\begin{eqnarray}
\bar{A}_{1}(\bar{j},t) \simeq \left\{ \begin{array}{ccc}
\sqrt{\frac{\bar{t}}{\pi}}&,  \bar{t} \ll 1&, \bar{j} > 0 \\
\rule{0mm}{3ex}\bar{t}&, \bar{t} \gg 1 \end{array} \right.
\end{eqnarray}
Note that the limits $\bar{t} \rightarrow 0$, $\bar{j} \rightarrow 0$ do not
commute. Taking $\bar{j} \rightarrow 0$ first, we from Eq.\ (4.7) find 
$2 \sqrt{\frac{\bar{t}}{\pi}}$, i.e.,
twice the result of the first line in Eq.\ (4.11). Below we will see that this
induces additional structure in the $t^{1/4}$-regime. Eqs.\ (4.8), (4.11)
yield
\begin{eqnarray}
\bar{g}_{1}(\bar{j},\bar{t}) = 2 \ell_{s} \sqrt{\frac{\rho_{0}}{\pi}}\:\cdot
\left\{ \begin{array}{cc}
\left(\frac{\bar{t}}{\pi}\right)^{1/4}&, \bar{t} \ll 1 \: \: \:,\\
\rule{0mm}{3ex}\bar{t}^{\,1/2}&, \bar{t} \gg 1 \end{array} \right.
\end{eqnarray}
equivalent to the well known asymptotic results
\begin{eqnarray}
g_{1}(j,N,t) \simeq \frac{2}{\sqrt{\pi}}\:\cdot \left\{ \begin{array}{cc}
\left(\frac{\tilde{t}}{\pi}\right)^{1/4}&, \: \: \: \: \: \bar{t}
\ll 1\: \:. \\
\rule{0mm}{3ex}\tilde{N}^{-1/2} \:\tilde{t}^{1/2}&, \: \: \: \: \: \bar{t}
\gg 1 \end{array} \right.
\end{eqnarray}
For any $\bar{j} > 0$,
the limit $\bar{t} \rightarrow 0$ is independent of $N$, since
for $t \ll T_{2} = O(N^{2})$ bead $j = \bar{j} N$ does not feel the presence
of the chain ends. Its motion is driven by the particles which initially sit
in its neighbourhood on the chain. In contrast, for $t \gg T_{2}$, the
motion is dominated by particles diffusing over the total length of the chain, which yields the factor $N^{-1/2}$, again independent of the position
of bead $j$ inside the chain.

The crossover among the limits of small or large time depends on $\bar{j}$.
Naively we might expect it to be governed by the time needed for particle diffusion
from the nearest reservoir to bead $j$:
\begin{equation}
\bar{T}_{2}(\bar{j}) \approx \bar{j}^{2}, \: \: \: \: \: \: 0 < \bar{j} \leq
\frac{1}{2} \: \: \:.
\end{equation}
For $0 < \bar{j} \ll \frac{1}{2}$ the situation, however, is even richer. The
crossover $t^{1/4} \leftrightarrow t^{1/2}$
always occurs at
\begin{eqnarray*}
\bar{t} \approx \bar{T}_{2} \left(\frac{1}{2}\right) \approx O (1) \: \: \:.
\end{eqnarray*}
It is due to particles diffusing over the whole length of the chain with nonnegligible probability.
For such times several terms nonnegligibly contribute to the sums in Eqs.\
(4.6), (4.7). For $\bar{j} \ll \frac{1}{2}, \bar{T}_{2}(\bar{j}) \alt \bar{t}
\ll \bar{T}_{2} \left(\frac{1}{2}\right)$, however, the $\nu = 0$
contribution of the second term in Eq.\ (4.7) is of the same order of magnitude as the $\nu = 0$ contribution of the first term, both being much
larger than all $\nu \neq 0$ terms. Thus the $t^{1/4}$ regime splits according to
\begin{eqnarray}
\bar{g}_{1}(\bar{j},\bar{t}) = 2 \ell_{s} \sqrt{\frac{\rho_{0}}{\pi}} \left(\frac{\bar{t}}{\pi}\right)^{1/4} \cdot \left\{ \begin{array}{cc}
1&, \: \: \: \: \: \bar{t} \ll \bar{T}_{2}(\bar{j})  \\
\sqrt{2}&, \: \: \: \: \: \bar{T}_{2}(\bar{j})  \ll \bar{t} \ll \bar{T}_{2}
\left(\frac{1}{2}\right)  \end{array} \right.
\end{eqnarray}
The crossover at $\bar{T}_{2}(\bar{j})$ is due to the particles diffusing in from the nearest chain end, while the
other chain end effectively is still infinitely far away.

The sums in Eqs.\ (4.6), (4.7) are rapidly converging in overlapping domains
of $\bar{t}$. $\bar{g}_{1}(\bar{j},\bar{t})$ therefore is easily evaluated
numerically. Fig.\ 3 shows typical results, normalized to the asymptotic
$\bar{t} \rightarrow 0$ behavior. The crossover from $\bar{g}_{1} \sim \bar{t}^{1/4}$ to $\bar{g}_{1} \sim \bar{t}^{1/2}$ occurs for $0.1 \alt \bar{t} \alt 1$. For $\bar{j} = 0.01$ the splitting of the
$\bar{t}^{1/4}$-regime is clearly seen.

\item[B)] We next consider the limit of long chains $N \rightarrow \infty$, intermediate beads $0 < \bar{j} < 1/2$ and shorter times $0 < t < \infty$.
In this limit for beads $j = \bar{j} N \gg 1$ both chain ends effectively are infinitely far away, and the
result becomes independent of $\bar{j}$. We thus describe the behavior of
some bead well inside the chain for times $t \ll T_{2} \approx O(N^{2}) \rightarrow \infty$. The analysis, based on Eq.\ (3.12), reveals that $A_{1}(j,t)$ reduces to (cf.\ Eq.\ (A.5))
\begin{equation}
A_{1}(j,t) = \frac{4}{\pi}\:p t\:\int_{0}^{\pi/2}\:dx\:\cos^{2} x (1 - 4
p\:\sin^{2} x)^{t-1}\: \: \:,
\end{equation}
with the asymptotic limit
\begin{equation}
A_{1}(j,t) \simeq \left(\frac{p t}{\pi}\right)^{1/2}, \: \: \: t \gg 1 \: \: \:.
\end{equation}
In this limit we recover the results found above for $\bar{t} \ll 1$.

The dominant corrections to that behavior arise from the contribution $F_{1}(4 \rho_{0} A_{1}(j,t))$ in Eq.\ (4.1). With Eqs.\ (4.17), (3.28) we
find
\begin{eqnarray}
g_{1}(j,N,t) = 2 \ell_{s} \left(\frac{\rho_{0}}{\pi}\right)^{1/2}
\left(\frac{p t}{\pi}\right)^{1/4} \left[ 1 - \frac{1}{16 \rho_{0}} \left(\frac{\pi}{p t}\right)^{1/2} + O \left(\frac{1}{t}\right)\right] \: \:
\:.
\end{eqnarray}
With $\rho_{0} \approx 1/4$ the leading order correction in (4.18) falls below the level of
1\% only for $p t \agt 5 \cdot 10^{3}$. Fig.\ 4 (full line) shows our full
result, normalized to the asymptotic $t^{1/4}$ behavior. We choose values
$\ell_{s} = 2,p = \frac{1}{5}, \rho_{0} = \frac{1}{4}$, but we should note
that for given $\ell_{s}$ the result essentially depends on the scaled variable $\rho_{0}^{2} p t \sim \tilde{t}$, only. The residual
$\rho_{0}$-dependence is negligible for $\rho_{0}^{2} p t \agt 1$. This reflects the fact that $A_{1}$ (Eq.\ 4.16) rapidly attains its asymptotic
behavior (4.17) the corrections dropping to the level of 1 \% for $p t \agt
10$. By virtue of Eq.\ (4.2) this implies that
$\left(\overline{n^{2}(j,t)}\right)^{1/2}$ rapidly tends to its asymptotic
$t^{1/4}$-behavior. The variation seen in Fig.\ 4 solely is due to the correction function $F_{1}(z)$, that reflects the discreteness 
of $n(j,t)$, cf. discussion after Eq.\ (3.26).

This long initial transient, extending almost up to $p t = O (10^{4})$, at a
first glance is quite surprising. A second thought reveals that it is quite
natural. With the estimate $|\overline{n(j,t)}| \sim (p t)^{1/4}$ we find
that in the region of interest typical values of $n(j,t)$ are of the order
$10$ or smaller. It is clear that with such small values of $n(j,t)$ the
discreteness of the process may induce large corrections.
\end{itemize}
We now also might discuss the limit $N \rightarrow \infty$ and $j,t$ fixed, implying that the bead considered has a finite distance from a chain end.
Here, however, tube renewal rapidly comes into play, and we therefore postpone the analysis to the next section. We close this discussion by illustrating the motion of the central bead $j = N/2$ for times $t \ll T_{3}$
in a chain of finite length. The results follow from numerical evaluation of
Eqs.\ (4.1), (4.2), with Eqs.\ (3.27), (4.4), (4.5) inserted. Fig.\ 4 (broken
lines) shows the typical behavior of $g_{1}\left(\frac{N}{2},N,t\right)$,
again normalized to the intermediate asymptotic $t^{1/4}$-behavior. For $N =
10^{3}$ this behavior is attained over almost three decades, until for $pt
\agt 10^{5}$ the curve bends over towards the $t^{1/2}$-regime. With decreasing $N$ the length of the $t^{1/4}$-plateau decreases rapidly, and is just about
visible for $N = 10^{2}$. For shorter chains it is washed out. In a
doubly logarithmic plot the result for $N=40$, for instance, for short times
would look like some effective power law close to $g_{1} \sim t^{1/3}$.

As shown in Fig.\ 3 the crossover towards the $t^{1/2}$-law starts for $p t
\approx N^{2}/10$ and takes more than one decade in $t$. The
$t^{1/2}$-behavior
reasonably well is approached for $p t \agt 2 N^{2}$. In the next section we
will find an estimate for the reptation time: $p T_{3} \approx
\frac{\pi}{16}\:N^{3}$. Thus the condition for the occurence of a true $t^{1/2}$-regime: $T_{2} \ll t \ll T_{3}$ takes the more quantitative form $2
N^{2} \alt p t \ll \frac{\pi}{16}\:N^{3}$. Clearly for typical chain lengths
used in simulations $(N < 10^{3})$ this condition is not fulfilled in a sufficiently large range of $p t$ to allow for an intermediate
$t^{1/2}$-region.

Above we stressed that the initial slow transient in $g_{1}(j,N,t)$ is due
to the correction $F_{1}(\ldots)$. $\left(\overline{n^{2}(j,t)}\right)^{1/2}$
accurately shows $t^{1/4}$-behavior starting as early as $\rho_{0}^{2}\:p t
= O(1)$. This quantity therefore shows standard reptation behavior much clearer than $g_{1}(j,N,t)$, and it is of interest to ask whether it can be
related to some simple observable. Indeed it is not hard to show that for the
lattice model described in Sect.\ 2, it is related to the cubic invariant
\begin{eqnarray*}
\hat{g}_{1}(j,N,t) &=& \left\langle \sum^{3}_{\alpha = 1}\:({\rm
r}_{j,\alpha}(t) - {\rm r}_{j,\alpha}(0))^{4} \right\rangle^{1/2}
\\
 &=& \ell_{s}\left(\overline{n^{2}(j,t)}\right)^{1/2}, \: \: \: \: \: t <
T_{R}(j) \: \: \:.
\end{eqnarray*}
This result holds in three dimensions. For other dimensions or other microstructures $\left(\overline{n^{2}(j,t)}\right)^{1/2}$ always can be
measured as an appropriate combination of the fourth and the second moment of
$({\bf r}_{j}(t) - {\bf r}_{j}(0))$.

\section{Tube renewal}

Tube renewal is just another word for the motion of the end beads that for longer times also affects the inner beads and the chain as a whole. As shown in Sect.\ II.A for chain end $0$ and times $t < T_{3}$ it leads us to consider $n_{max}(t)$, defined as the
maximal value
within time interval $[0,t]$ of the stochastic variable $n(0,t)$, that gives
the change in the occupation number of reservoir $0$ (cf.\ Eq.\ (2.10)). In
particular the relation (2.12)
\begin{eqnarray*}
g_{1}(0,N,t) &=& \left\langle \overline{({\bf r}_{0}(t) - {\bf
r}_{0}(0))^{2}} \right\rangle
\\
&=& 2 \ell_{s} \overline{n_{max}(t)}
\end{eqnarray*}
holds.\\
As is obvious from Eq.\ (3.13), $n(0,s)$ executes a correlated random walk with $n (0,0) = 0$,
the correlation arising from the contribution $\sim A_{2} (s_{2} - s_{1})$: a
particle, that jumped out of the reservoir at time $s_{1}$, with some probability depending on $s_{2} - s_{1} > 0$ will fall back again at time
$s_{2}$. For correlated processes no manegeable exact expression for $\overline{n_{max}(t)}$ is known. Only for uncorrelated walks a result known
as Spitzer's identity \cite{Z17} yields a simple answer. (This identity is
further discussed in appendix B.) For a reflection symmetric uncorrelated
walk $n_{s}$ starting at $n_{0} = 0$, it yields the relation
\begin{equation}
\overline{n_{max}(t)} = \sum^{t}_{s=1}\: \frac{\overline{|n_{s}|}}{2s}
\end{equation}
    We will base our analysis on this result. To get some feeling for the
structure of the problem and the approximation involved we, however, first
derive a formally exact expression for $\overline{n_{max}(t)}$. We then approximately treat the motion of a chain end, and with a further approximation we calculate the motion of some bead $j$ for times $t > T_{R}(j)$. Finally we discuss the reptation time. 

\subsection{Formal analysis of $\overline{n_{max}(t)}$}
From Eqs.\ (3.2), (3.17), (3.13) the distribution of $\{n_{s}\} \equiv \{n(0,s)\}$ is found as
\begin{equation}
\hat{{\cal P}}_{2}(\{n_{s}\};t) = (2\pi
i)^{-t}\:\oint\:\prod^{t}_{s=1}\:\frac{d x_{s}}{x_{s}}\:x_{s}^{-n_{s}}\:\exp
Q_{2} (\{x_{s}\};t) \: \: \:,
\end{equation}
where all integrals can be taken to range over the unit circle. It is useful
to transform from variables $x_{s}$ conjugate to $n_{s}$ to variables $y_{s}$ conjugate to the increments $n_{s}-n_{s-1}$.
\begin{equation}
x_{s} = \frac{y_{s+1}}{y_{s}}, \: \: \: \: \: 1 \leq s \leq t;\: \:  y_{t+1}
\equiv 1
\end{equation}
Successive substitution yields
\begin{equation}
\hat{{\cal P}}_{2}(\{n_{s}\};t) = (2\pi
i)^{-t}\:\oint\:\prod^{t}_{s=1}\:\frac{ dy_{s}}{y_{s}}\:\left(
\frac{y_{s}}{y_{s+1}}\right)^{n_{s}} \:\exp \tilde{Q}_{2}
(\{y_{s}\};t) \: \: \:,
\end{equation}
where
\begin{eqnarray}
\tilde{Q}_{2}(\{y_{s}\};t) &=& \rho_{0}\:p\:\sum^{t}_{s=1}\:\left(y_{s} +
\frac{1}{y_{s}} - 2\right)
\nonumber \\
& & + \rho_{0} p^{2}\:\sum_{0<s_{1}\leq s_{2}<t}\:\left(\frac{1}{y_{s_{1}}} -
1\right) \left(y_{s_{2} +1} - 1\right) A_{2} (s_{2} - s_{1})
\end{eqnarray}

To derive an expression for $n_{max}(t)$ we define the operator
\begin{equation}
B(n,t) = \prod^{t}_{s=0}\:\Theta (n - n_{s}) \: \: \:,
\end{equation}
with the discrete $\Theta$ function as defined in (2.14).
$B(n,t)$ obeys the relation
\begin{equation}
B (n,t) - B (n-1,t) = \delta_{nn_{max}(t)} \: \: \:.
\end{equation}
We thus have
\begin{equation}
\overline{n_{max}(t)} = \sum^{\infty}_{n=1}\:n\:\left(\overline{B(n,t)} -
\overline{B(n-1,t)}\right) \: \: \:.
\end{equation}
We next introduce
\begin{equation}
{\cal C} (t) = \overline{n_{max}(t)} - \overline{n_{max}(t-1)} \: \: \:,
\end{equation}
so that
\begin{equation}
\overline{n_{max}(t)} = \sum^{t}_{s=1}\:{\cal C}(s) \: \: \:.
\end{equation}
Combining
Eqs.\ (5.8), (5.9) we easily find
\begin{eqnarray}
{\cal C}(t) &=& - \sum^{\infty}_{n=1}\: \left(\overline{B(n,t)} -
\overline{B(n,t-1)}\right)
\nonumber \\
&=& \sum^{\infty}_{n=0}\:\sum^{n}_{n_{1}\ldots n_{t-1} = -
\infty}\:\sum^{\infty}_{n_{t}=n+1}\: \hat{{\cal P}}_{2} (\{n_{s}\};t) \: \:
\:.
\end{eqnarray}
In Eq.\ (5.4) for $\hat{{\cal P}}_{2} (\{n_{s}\};t)$ we now order the integration contour ${\cal B}$ as
\begin{equation}
{\cal B}: 1 > |y_{1}| > |y_{2}| > \ldots > |y_{t}|
\end{equation}
We then in Eq.\ (5.11) can carry through all summations to find
\begin{equation}
{\cal C}(t) = (2 \pi i)^{-t}\:\oint_{{\cal B}}\:\prod^{t}_{s=1}\:\frac{d
y_{s}}{y_{s} - y_{s+1}}\:\frac{1}{y_{1}-1}\:\exp 
\tilde{Q}_{2}(\{y_{s}\};t)
\end{equation}
Recall the definition $y_{t+1} \equiv 1$.

Eq.\ (5.13) is an exact but essentially intractable expression. To establish
the connection to Eq.\ (5.1) we note the identity
\begin{equation}
(2 \pi i)^{-t}\:\oint_{{\cal B}}\:\prod^{t}_{s=1}\:\frac{d y_{s}}{y_{s} -
y_{s+1}}\:\frac{1}{y_{1}-1}\:\sum^{t-1}_{s'=0}\:{\cal S}_{s'} F\{y_{s}\}
- \frac{1}{2\pi i}\:\oint_{b<|y|<1}\:dy\:\frac{F\{y,\ldots,y\}}{(y-1)^{2}} =
0 \: \: \:,
\end{equation}
where ${\cal S}_{s'}$ is a cyclic shift operator:
\begin{equation}
{\cal S}_{s'} F\{y_{1},\ldots,y_{t}\} = F
\{y_{1+s'},y_{2+s'},\ldots,y_{t},y_{1},y_{2},\ldots,y_{s'}\}\:\:\:.
\end{equation}
This identity holds for any function $F\{\ldots\}$ analytic in all variables
$y_{s}$ in the domain
\begin{eqnarray*}
a > |y_{s}| > b\: \:; \: \: \: a > 1 > |y_{t}| > b
\end{eqnarray*}
We now note that
\begin{eqnarray*}
\tilde{Q}_{2}(\{y,\ldots,y\};t) \equiv Q_{1} (y;0,t) \: \: \:,
\end{eqnarray*}
which yields
\begin{equation}
\frac{1}{2\pi
i}\:\oint_{|y|<1}\:dy\:\frac{\exp \tilde{Q}_{2}(\{y,\ldots,y;t)}{(y-1)^{2}} 
= \frac{1}{2}\: \overline{|n(0,t)|}
\end{equation}
(cf.\ Eq.\ 3.25). We then subtract from $C(t)$ (Eq.\ 5.13) the identity (5.14) multiplied by $1/t$, to find
\begin{eqnarray}
{\cal C}(t) &=& \frac{\overline{|n(0,t)|}}{2 t}
\nonumber \\
& & + (2 \pi i)^{-t}\:\oint_{{\cal B}}\:\prod^{t}_{s=1}\:\frac{d y_{s}}{y_{s}
- y_{s+1}}\:\frac{1}{y_{1}-1}\:\left(1 - \frac{1}{t} \sum^{t-1}_{s'=0}\:{\cal
S}_{s'}\right) \exp \tilde{Q}_{2}(\{y_{s}\};t)
\end{eqnarray}
We note that the contribution $|\overline{n(0,t)}|/2 t$ takes into account
the strongest singularity of the integrand in Eq.\ (5.13), occuring if all
variables $y_{s}$ coincide.

For an uncorrelated random walk the exponential factorizes: $\exp
\tilde{Q}_{2}(\{y_{s}\},t) \rightarrow \prod^{t}_{s=1}\:f(y_{s})$, and the
second contribution in Eq.\ (5.17) vanishes. In view of Eq.\ (5.10) the result therefore yields a formal generalization of Spitzer's identity (5.1)
to correlated walks.
Transformed back to the variables $n_{s}$, the second term in Eq.\ (5.17)
weights each walk contributing to ${\cal C}(t)$ (Eq.\ 5.11) by the difference
of its proper weight and the average weight of all of its cyclic
permutations. Depending on the walk considered, the result can be of either
sign, and we may hope that cancellations keep the contribution of the second
term in Eq.\ (5.17) small as compared to the first term. We, however, have found
no proper estimate.

More can be said only in the limit $N \rightarrow \infty, \bar{t} = p t/N^{2}$ fixed (case A of Sect.\ 4). For $\bar{s} = p s/N^{2} \gg 1$ the
coupling term $A_{2}(s)$ in $Q_{2}$ (Eq.\ (3.13)) vanishes exponentially,
reflecting the fact that over time intervals of order $T_{2} \sim N^{2}$ the
particle distribution within the chain equilibrates. A process with coarse
grained time intervals $\tau \gg T_{2}$ proceeds by steps of average size
$(\Delta n)^{2} \sim \tau \gg 1$, with correlations of order $1$ 
of a time step with its immediate precursor. Neglecting these small
correlations we find that Eq.\ (5.1) gives the leading contribution to $\overline{n_{max}(t)}$ for $\bar{t} \rightarrow \infty$. In this long time
regime chain motions can be modelled as an uncorrelated random walk of a
single degree of freedom \cite{Z1,Z14}.

\subsection{Motion of the end bead in the `mean hopping rate' approximation}

We here evaluate $g_{1}(0,N,t)$ as
\begin{equation}
g_{1}(0,N,t) = \ell_{s}\:\sum^{t}_{s=1}\:\frac{1}{s}\:\overline{|n(0,s)|} \: \: \:,
\end{equation}
(cf.\ Eqs.\ (2.12), (5.1)), with
\begin{equation}
\overline{|n(0,s)|} = \frac{2}{\sqrt{\pi}}\:\left(\rho_{0} A_{1}(0,s)\right)^{1/2}
\left[ 1 - F_{1} (4 \rho_{0} A_{1} (0,s))\right] \: \: \:,
\end{equation}
(cf.\ Eq.\ (3.26)).

Since in Eq.\ (5.18) the contribution ${\cal C}(s)$ of a time step $s$ is
calculated as the contribution of an uncorrelated walk giving the same (time dependent) distribution of $n(0,s)$ as the true correlated walk, we call this a `mean
hopping rate' approximation. For the interpretation of our results it is
useful to determine the effective hopping rate $p_{eff}(s)$ of this uncorrelated
walk $n_{eff}(s)$. Writing
\begin{eqnarray*}
\overline{n^{2}_{eff}(s)} = 2 p_{eff}\:s
\end{eqnarray*}
and identifying $\overline{n^{2}_{eff}(s)}$ and $\overline{n^{2}(0,s)}$ we find
\begin{equation}
p_{eff}(s) = \frac{\rho_{0}}{s}\:A_{1}(0,s) \: \: \:.
\end{equation}
Eq.\ (3.12) yields $A_{1}(0,1) = p$, so that the expected result
\begin{equation}
\eqnum{5.21i}
p_{eff}(s = 1) = \rho_{0}\:p
\end{equation}
follows. For large $s$, however, the first term in Eq.\ (3.12) dominates and
we find
\begin{equation}
\eqnum{5.21ii}
p_{eff} (s \rightarrow \infty) \rightarrow \frac{\rho_{0} p}{N}
\end{equation}
\setcounter{equation}{21}
Thus the effective hopping rate strongly decreases with increasing time,
reflecting the fact that particles emitted from the reservoir may have been absorbed
again.

We now first consider the motion of the end bead (5.18), (5.19) in several limits, corresponding to those of Section IV for some internal bead.
\begin{itemize}
\item[A)] $N \rightarrow \infty, \bar{t}$ fixed. \\
Since $\overline{|n(0,s)|}$ increases with $s$, the sum (5.18) is dominated
by large values of $s$. In the limit considered, we therefore can neglect the
correction $F_{1}(\ldots)$, and we can replace the summation by an integral.
This yields
\begin{eqnarray}
\bar{g}_{1}(0,\bar{t}) &=& \lim_{{\scriptsize\begin{array}{cc} N
&\rightarrow \infty \\ \bar{t}& \mbox{fixed} \end{array}} }\:
N^{-1/2} g_{1}(0,N,t)
\nonumber \\
&=& 2 \ell_{s}\:\sqrt{\frac{\rho_{0}}{\pi}}\:\int_{0}^{\bar{t}}\:\frac{d
\bar{s}}{\bar{s}}\:\bar{A}_{1}^{1/2} (0,\bar{s}) \: \: \:,
\end{eqnarray}
where
\begin{equation}
\bar{A}_{1}(0,\bar{s}) \simeq \left\{ \begin{array}{cc}
2\sqrt{\frac{\bar{s}}{\pi}} &, \: \: \: \: \: \bar{s} \ll 1 \\
\rule{0mm}{3ex}\bar{s} &, \: \: \:\bar{s} \gg 1 .
\end{array} \right.
\end{equation}
(Cf.\ Eq.\ (4.11) and the associated discussion.)

For $\bar{t} \gg 1$ we find
\begin{equation}
\bar{g}_{1}(0,\bar{t}) = 4 \ell_{s} \left(\frac{\rho_{0}
\bar{t}}{\pi}\right)^{1/2} \: \: \:,
\end{equation}
a result which in the light of the discussion of the previous section is
expected to be exact. Comparing to the motion of an inner segment $\bar{j} > 0$, Eq.\ (4.12), we find
\begin{equation}
\frac{\bar{g}_{1}(0,\bar{t})}{\bar{g}_{1}(\bar{j},\bar{t})}\:
\stackrel{\bar{t} \rightarrow \infty}{=}
 \: 2 \: \: \:.
\end{equation}
Recall that due to tube renewal the effective random walk of an end bead
in the limit considered has twice the length of the walk of an interior bead, cf. Eqs.\ (2.8) and (2.12) with $n_{max}(t) \simeq |n(j,t)|$. This is the origin of the result (5.25).

For $\bar{t} \ll 1$ Eqs.\ (5.22), (5.23) yield
\begin{equation}
\bar{g}_{1}(0,\bar{t}) = 8 \sqrt{2}\:\ell_{s}
\left(\frac{\rho_{0}}{\pi}\right)^{1/2}
\left(\frac{\bar{t}}{\pi}\right)^{1/4}
\: \: \:,
\end{equation}
or
\begin{equation}
\frac{\bar{g}_{1}(0,\bar{t})}{\bar{g}_{1}(\bar{j},\bar{t})}\:
\stackrel{\bar{t} \rightarrow 0}{=}
 \: 4\: \sqrt{2} \: \: \:.
\end{equation}
This result is not trivial. Naively we would estimate the ratio (5.27) as
$2\:\sqrt{2}$, the factor 2 reflecting tube renewal and the factor $\sqrt{2}$
arising from the enhanced mobility of the end bead. (Cf.\ the Poisson representation (4.7) for $\bar{t} \rightarrow 0, \bar{j} = 0$, as compared to
$\bar{j} > 0$.) The additional factor of 2 found in Eq.\ (5.27) is due to
the integration over $\bar{s}$ (Eq.\ 5.22), reflecting the fact that $n_{max}$ depends on the history of the walk. In our approximation this enhancement of the ratio (5.27) can be seen as a direct consequence of the
slow decrease of the mean hopping rate discussed above.

Fig.\ 5 shows the crossover behavior of the ratio
$\bar{g}_{1}(0,\bar{t})/\bar{g}_{1}\left(\frac{1}{2},\bar{t}\right)$. The
crossover from the asymptotics (5.27) to (5.25) sets in at $\bar{t} \approx
0.1$, i.e., $t \approx T_{2}/10$. It is quite slow, taking about three decades of $t$. We should note that at time $t \approx T_{3}$, formally infinite in the present limit, there must be another crossover towards $g_{1}(0,N,t)/g_{1}\left(\frac{N}{2},N,t\right
) = 1$, since eventually the chain is displaced as a whole.
\item[B)] $N \rightarrow \infty,t$ fixed \\
This limit shows the approach towards the asymptotic law (5.26), now adequately written as
\begin{equation}
g_{1}(0,\infty,t)
\stackrel{t \gg 1}{=}
 \: 8\: \sqrt{2} \ell_{s}
\left(\frac{\rho_{0}}{\pi}\right)^{1/2} \left(\frac{p t}{\pi}\right)^{1/4}
\: \: \:.
\end{equation}
In the general expressions no special simplifications occur, except that
$A_{1}(0,t)$ (Eq.\ (3.12)) takes the form
\begin{eqnarray}
A_{1}(0,t) &=& 8 p t\:\int_{0}^{1/2}\:dx\:\cos^{2} (\pi x) \left(1 - 4 p\:\sin^{2}(\pi x)\right)^{t-1}
\nonumber \\
& & + \int_{0}^{1/2}\:dx\:\left(1 - 4 p \:\sin^{2}(\pi x)\right)^{t} - \frac{1}{2} \: \: \:.
\end{eqnarray}
The result for $g_{1}(0,\infty,t)$, normalized to the asymptotics (5.28), is
shown as full line in Fig.\ 6. As compared to the behavior of an interior
segment (Fig.\ 4) we find a drastic enhancement of the initial effects. (Note
the change in the scales.) The $t^{1/4}$-regime is reached only for $p t
\geq 10^{8}$. This again is due to the discrete nature of the process, which
here comes in at two places: Firstly, $|n(0,s)|$ involves the correction
$F_{1} (4 \rho_{0} A_{1}(0,t))$ (cf.\ Eq.\ 5.19), and secondly, $n_{max}(t)$
involves the {\em sum} over all time steps. The long initial transient is
dominated by the latter feature and thus quite insensitive to the value of
$\rho_{0}$, that only influences $F_{1}(\ldots)$.
\end{itemize}

Evaluating $g_{1}(0,N,t)$ for finite $N$ and normalizing by the intermediate
asymptotics we find results as shown by the broken curves in Fig.\ 6 . For
the end bead it needs extremely long chains to see indications of $t^{1/4}$-behavior. For chain lengths $N \alt 100$, typically used in computer experiments, the initial transient directly bends over towards free
diffusion. Clearly the observation of an intermediate $t^{1/2}$-regime is
beyond all reach.

\subsection{Motion of an arbitrary bead}

As shown by Eq.\ (2.13), the motion of some bead $0 < j < N$ for times $t <
T_{3}$ is a superposition of motion within the tube and tube renewal. This
leads to a rich structure, which we first discuss on a qualitative level. A
schematic plot, summarizing our discussion for some bead $1 \ll j \ll N/2$, is shown in Fig.\ 7.

Consider some bead $1 \ll j \ll \frac{N}{2}$ in an extremely long chain. Initially it
does not feel the chain ends but shows the slow transient towards a $t^{1/4}$-behavior, as discussed in Sect.\ 4:
\begin{equation}
g_{1}(j,N,t) \rightarrow
2\:\ell_{s}\:\sqrt{\frac{\rho_{0}}{\pi}}\:\left(\frac{p t}{\pi}\right)^{1/4}
, \: \: \: T_{0} \ll t \ll T_{2}(j), \: \: \: p T_{2}(j) \approx j^{2} \: \: \:.
\end{equation}
For $t \approx T_{2}(j)$, particles coming from the nearest chain end start
to diffuse over bead $j$, which raises the amplitude of the $t^{1/4}$-law by
a factor of $\sqrt{2}$, cf. Eq.\ (4.15), thus
\begin{equation}
g_{1} (j,N,t) \approx 2 \sqrt{2}\:\ell_{s}
\sqrt{\frac{\rho_{0}}{\pi}}\:\left(\frac{p t}{\pi}\right)^{1/4}, \: \: T_{2}(j) \ll t \ll T_{R}(j)
\end{equation}
This regime will end at a time $T_{R}(j)$ defined as the time needed for
the tube renewal to pass over bead $j$:
\begin{equation}\
\ell_{s}\:\overline{n_{max}(T_{R}(j))} = j
\end{equation}
On the basis of Eq.\ (5.28) $T_{R}(j)$ is estimated as
\begin{equation}
p T_{R}(j) \approx \frac{1}{2} \left(\frac{\pi}{8}\right)^{3}
\frac{j^{4}}{(\ell_{s}^{2} \rho_{0})^{2}},
\end{equation}
For $t > T_{R}(j)$, tube renewal dominates. For extremely long chains it
yields another $t^{1/4}$-regime.
\begin{equation}
g_{1} (j,N,t) \approx 8 \sqrt{2}\:\ell_{s}\:
\sqrt{\frac{\rho_{0}}{\pi}}\:\left(\frac{p t}{\pi}\right)^{1/4}, \: \: T_{R}(j) \ll t \ll T_{2}(N)
\end{equation}
Gradually the $t^{1/2}$-behavior sets in beyond the Rouse time $T_{2}(N)$:
\begin{equation}
g_{1} (j,N,t) \approx 4 \:\ell_{s}\: \left( \frac{\rho_{0}}{\pi} \frac{p
t}{N}\right)^{1/2},  \: \: T_{2}(N) \ll t \ll T_{3}\: \: \:,
\end{equation}
and finally we end up with free diffusion
\begin{equation}
g_{1} (j,N,t) \approx\: const\:\frac{t}{N^{2}}, \: \: \: \: \: T_{3} \ll t
\end{equation}
$T_{3}$ may be identified with $T_{R} (\frac{N}{2})$, calculated from Eq.\
(5.32). For $j = O(\frac{N}{2})$ we, however, can not use Eq.\ (5.33). In
evaluating Eq.\ (5.32) we rather have to resort to
Eq.\ (5.24) to find
\begin{equation}
p T_{R}(j) \approx \frac{\pi}{4}\:\frac{j^{2} N}{\ell_{s}^{2} \rho_{0}} \:
\: \:,  j = O (\frac{N}{2})
\end{equation}

The above analysis breaks down for $T_{R}(j) \agt T_{2}(N)$. With $T_{2}(N)
\sim N^{2}, T_{R}(j) \sim j^{4}$ we find it to be valid only for $1 \ll j \ll
O (N^{1/2})$. For larger values of $j$ the $t^{1/4}$-regime (5.34), 
driven by the motion of the end bead, vanishes. Then tube renewal under
favourable conditions may introduce some structure in the $t^{1/2}$-regime.
We, however, should note that this whole discussion is more a matter of principle. It will need extremely long chains $(N \agt 10^{4})$ to observe
these different regions.
In practice we expect that only the initial $t^{1/4}$ regime (5.30) properly
can be observed, the other regimes being covered by broad crossover regions.

For a more quantitative analysis of $g_{1}(j,N,t)$ as given by Eq.\ (2.13)
we have to construct the simultaneous distribution of $n_{max}(t)$ and $n(j,t)$.
\begin{equation}
\hat{{\cal P}}_{max,j}(n_{m},n;t) = \overline{\delta_{n_{max}(t),n_{m}} \delta_{n(j,t),n}}
\end{equation}
Again the formal expression resulting from Eqs.\ (3.11) - (3.17) is intractable. To construct an approximate form we introduce $n_{t} \equiv
n(0,t)$ as intermediate variable and write
\begin{eqnarray}
\hat{{\cal P}}_{max,j}(n_{m},n;t) &=& \sum^{+ \infty}_{n_{t} = - \infty}\:
\overline{\delta_{n_{max},n_{m}} \delta_{n(0,t),n_{t}} \delta_{n(j,t),n}}
\nonumber \\
&=& \sum^{+ \infty}_{n_{t} = - \infty}\:\hat{{\cal
P}}_{max,0,j}(n_{m},n_{t},n;t) \: \: \:.
\end{eqnarray}
We then factorize according to
\begin{equation}
\hat{{\cal P}}_{max,0,j}(n_{m},n_{t},n;t) \rightarrow \hat{{\cal P}}_{max,0}
(n_{m} | n_{t};t) \hat{{\cal P}}_{2} (n_{t},n;0,j,t) \: \: \:,
\end{equation}
where $\hat{{\cal P}}_{2} (n_{t},n;0,j,t)$ is the simultaneous distribution
of $n(0,t),n(j,t)$ discussed in Sect.\ 3, and $\hat{{\cal P}}_{max,0}
(n_{m} | n_{t};t)$ is the conditional probability to find $n_{max}(t) = n_{m}$, given $n(0,t) = n_{t}$. This approximation serves to isolate the hard
part of the problem in the analysis of the single stochastic variable $n(0,s)$. It is justified by the observation that for beads not too close to
a chain end $(j \agt 20$, say), the relation $T_{R}(j) \gg T_{2}(j)$ holds,
so that bead $j$ feels the tube renewal only for times where $n(j,t)$ is
bound to $n(0,t)$ fairly rigidly. (Recall the discussion after Eq.\ (3.35).)

To find an approximate expression for the conditional probability
$\hat{{\cal P}}_{max,0}(n_{m} | n_{t};t)$ we resort to the simple random
walk, which is the only model where this probability can be calculated in
explicit closed form. We thus ignore all the correlations of the stochastic
process $n(0,s)$, but we keep the important inequalities $0 \leq n_{max}(t),
n(0,t) \leq n_{max}(t)$, and we can adjust the hopping probability of the
random walk so as to reach the desired values of $\overline{n_{max}(t)}$. As
a final purely technical point we take the variables $n_{m},n_{t},n$ to be of
continuous range, which is a minor approximation since for beads not too
close to a chain end we are interested only in fairly large times $(p t \agt
p T_{R}(j) \agt 10^{3})$.

With these approximations the explicit construction of $\hat{{\cal P}}_{max,j}(n_{m},n;t)$ is carried through in appendix C. The final result
reads
\begin{eqnarray}
\lefteqn{
\hat{{\cal P}}_{max,j}(n_{m},n;t)= \frac{\Theta
(z_{m})}{\overline{n_{max}(t)}
\left(\overline{n^{2}(j,t)}\right)^{1/2}}\:\frac{1}{\pi}}
\nonumber \\
& & \Bigg\{\sqrt{\frac{2}{\pi}}\:(1 - a^{2})^{1/2}
\exp \left[- \frac{1}{1-a^{2}} \left( \frac{z_{m}^{2}}{\pi}\: - a \sqrt{\frac{2}{\pi}}\:z_{m} z + \frac{z^{2}}{2}\right) \right]
\nonumber \\
& & + a \left( 2 a
\sqrt{\frac{2}{\pi}}\:z_{m} - z\right)
\exp \left[ - \frac{1}{2} \left(2 a \sqrt{\frac{2}{\pi}}\:z_{m} - z\right)^{2} \right] \:\mbox{erfc}\: \left( \frac{\sqrt{\frac{2}{\pi}}\:z_{m}(1 - 2
a^{2}) + a z}{\sqrt{2 (1 - a^{2}}} \right) \Bigg\} \: \:,
\end{eqnarray}
where
\begin{eqnarray*}
z_{m} = \frac{n_{m}}{\overline{n_{max}(t)}}, \: \: \: z =
\frac{n}{(\overline{n^{2}(j,t)})^{1/2}} \: \: \:,
\end{eqnarray*}
and
\begin{equation}
a = a(j,t) = \frac{\tilde{A}_{3}(j,t)}{(A_{1}(0,t) A_{1}(j,t))^{1/2}}
\end{equation}
$\tilde{A}_{3}(j,t)$ is defined in Eq.\ (3.31). The parameter $a, 0 \leq a
\leq 1$, measures the strength of the correlation among $n(0,t)$ and $n(j,t)$. For $a \rightarrow 0$, which is reached for $t \ll T_{2}(j)$, the
two processes are independent. They are tightly bound together for $a \rightarrow 1$, corresponding to $t \gg T_{2}(j)$.

To give an impression of the shape of the distribution we in Fig.\ 8 have
plotted ${\cal P}(z_{m},z,t) = \overline{n_{max}(t)} (\overline{n^{2}(j,t)})^{1/2}\:\hat{{\cal
P}}_{max,j}(n_{m},n;t)$ for fixed $z_{m} = 1,2,3$ as function of $z$, using
values $a = 0.9,0.99$. Specifically for $t = T_{R}(j)$, i.e., for times where
the tube renewal just reaches segment $j$, such values $a = a(j,T_{R}(j))$
correspond to $j = 50$ or $120$, respectively. As is shown, with increasing
$a$ the curves become strongly asymmetric. This reflects the distribution of
$n(0,t)$, which is cut off sharply at $n(0,t) = n_{max}(t)$. We also note
that tube renewal will have a substantial effect even for $t < T_{3}(j)$.
Both conditions $\ell_{s}\:n_{max}(t) > j, \ell_{s}\:n_{max}(t) > j + \ell_{s}\:n(j,t)$ then are fulfilled for $z_{m} > 1, z < 0$, a region that
carries a considerable fraction of the weight of the distribution $\hat{{\cal
P}}_{max,j}$.

We now evaluate $g_{1}(j,N,t)$, decomposing Eq.\ (2.13) as
\begin{equation}
g_{1}(j,N,t) = g_{i}(j,N,t) + g_{r}(j,N,t) \: \: \:,
\end{equation}
where
\begin{equation}
g_{i}(j,N,t) = \ell_{s}\:\overline{|n(j,t)|}
\end{equation}
is the contribution of the motion in the fixed initial tube, as discussed in
Sect.\ 4. The contribution of the tube renewal can be written as
\begin{eqnarray}
g_{r}(j,N,t) &=& 2 \ell_{s}\:\overline{\Theta (- n(j,t) - 1) \Theta \left(n_{max}(t) - \frac{j}{\ell_{s}}\right) \left(n_{max}(t) -
\frac{j}{\ell_{s}}\right)}
\nonumber \\
&+& 2 \ell_{s} \overline{\Theta (n(j,t)) \Theta \left(n_{max}(t) - n(j,t) -
\frac{j}{\ell_{s}}\right) \left(n_{max}(t) - n(j,t) -
\frac{j}{\ell_{s}}\right)}
\end{eqnarray}
With our appoximations it yields
\begin{eqnarray}
g_{r}(j,N,t) &=& 2 \ell_{s}\:\overline{|n_{max}(t)|}\:\int_{0}^{\infty}\:d
z_{m}\:z_{m} \cdot \Bigg\{G \left(z_{m} +
\frac{j}{\ell_{s}}\:\overline{|n_{max}(t)|}^{-1},0,a(0,j,t)\right)
\nonumber \\
& & + G \left(z_{m} + \frac{j}{\ell_{s}}\:\overline{|n_{max}(t)|}^{-1}, \frac{\left(\frac{2}{\pi}
\overline{n^{2}(j,t)}\right)^{1/2}}{\overline{|n_{max}(t)|}},\:a
(0,j,t)\right)\Bigg\}
\end{eqnarray}
\begin{eqnarray}
G (z,b,a) &=& \frac{1}{\pi}\:\left(1 - 2 a b + b^{2}\right)^{-1/2}\:\frac{1 -
ab}{1 - 2 a b}\:\exp\:\left[ - \frac{z^{2}}{\pi}\:\left(1 - 2 a b + b^{2}\right)^{-1}\right]
\nonumber \\
& & \cdot\:\mbox{erfc}\: \left(\frac{z}{\sqrt{\pi(1 - a^{2})}}\:\frac{ b - a}{(1 -
2 ab + b^{2})^{1/2}}\right)
\nonumber \\
& & - \frac{a}{\pi}\:(1 - 2 a b)^{-1}\:\exp\:\left[- 4 a^{2}
\frac{z^{2}}{\pi}\right]\:erfc\:\left(\frac{1 - 2 a^{2}}{\sqrt{\pi(1 - a^{2})}}\:z\right) \: \: \:.
\end{eqnarray}
Typical numerical results are shown in Fig.\ 9. Together with the full results for $N = 10^{3}, j = 20$ or $100$, we included $g_{i}(\ldots)$ and
$g_{r}(\ldots)$, and we show $g_{1}(0,N,t)$, for comparison. In
$g_{i}(\ldots)$ the first splitting of the $t^{1/4}$-regime can be seen both
for $j = 20$ and $j = 100$, but $g_{r}(\ldots)$ only for $j = 20$ shows weak
indications of the $t^{1/4}$-behavior from the end bead motion. For $j = 100$
tube renewal is effective only for $t \agt T_{2} \left(\frac{N}{2}\right)$.
In both cases we see that the total motion $g_{1}(j,N,t)$ beyond the first
$t^{1/4}$-regime is dominated by the crossover. We also note that tube renewal effects set in for $t \agt T_{R}(j)/10$, a consequence of the fairly
broad distribution $\hat{{\cal P}}_{max,j}(n_{m},n;t)$.

\subsection{An estimate of the reptation time}

The reptation time $T_{3}$ measures the time needed to destroy the original
tube. This process proceeds by the motion of both chain ends, and a simple
definition consists in identifying $T_{3}$ with $T_{R}(N/2)$, Eq.\ (5.31):
\begin{equation}
\ell_{s}\:\overline{n_{max}(T_{3})}\: = \frac{N}{2} \: \: \:.
\end{equation}
Other, more complicated definitions might be given, that take into account
the first-passage-time nature of the problem. The present definition closely
is related to the evaluation of the melt viscosity in the reptation model
\cite{Z14,Z15,Z18}. With Eq.\ (2.12) it implies
\begin{equation}
g_{1}(0,N,T_{3}) = N = R_{e}^{2}\: \: \:,
\end{equation}
where $ R_{e}^{2}$ is the mean squared end-to-end distance of the chain. We
should note, however, that strictly speaking for $t = T_{3}$ Eq.\ (2.12) is
invalid due to the interference of chain renewal from both ends, so that the
physical interpretation of $T_{3}$, as implied by Eq.\ (5.49), is only approximate.

For long chains we may invoke Eq.\ (5.37) to find the well known asymptotic
$N^{3}$-law:
\begin{equation}
p T_{3} (N)
\stackrel{N \rightarrow \infty}{\longrightarrow}
\:\frac{\pi}{16} \frac{N^{3}}{\ell_{s}^{2} \rho_{0}}
\end{equation}
To estimate the corrections we first may evaluate Eq.\ (5.48) with the form
(5.22) of $\ell_{s} \overline{n_{max}(t)} = N^{1/2}
\bar{g}_{1}(0,\bar{t})/2$, valid for long chains. This is equivalent to Doi's
model \cite{Z15} and yields the thin line in Fig.\ 10. As is seen this curve
bends downwards, yielding a strong increase in the effective exponent $T_{3}
\sim N^{x}$ with decreasing $N$. In fact, naively extending this curve towards $N \rightarrow 0$, we approach $x = 4$, reflecting the $t^{1/4}$-law.
Of course this extension is meaningless since $T_{3}(N)$ by construction is
far beyond the range where the $t^{1/4}$ law is valid. It is not surprising that the result is strongly modified if we take the discreteness
of the system into account. Our result using the full expression 5.1 for
$n_{max}(t)$ is shown as the fat line in Fig.\ 10. For $N \agt 200$ it merges
with the continuum approximation, but for the full range $20 \alt N  \alt
1300$ it is nicely approximated by $p T_{3}(N) \approx 0.059 N^{3.15}$. We
thus see that the slowing down of the increase of $n_{max}(t)$ with increasing $t$ implied by the correlations increases the effective exponent
beyond $x = 3$, but not as strongly as expected from a continuum
approximation.

Evaluating the Doi-Edwards expression for the melt viscosity $\eta$ we find
results for $\eta (N)$ most similar in all respects to those shown for $T_{3}(N)$ in Fig.\ 10. For $20 \alt N \alt 10^{3}$ the full theory yields an
effective power law $\eta(N) \sim N^{3.2}$, with the continuum approximation
bending downwards as in Fig.\ 10. Note that these results are quite similar
to those of \cite{Z18}, Fig.\ 3, where Doi's `Rouse chain in a tube' model
was simulated.

\section{Conclusions}

In this work we evaluated a discrete version of the reptation model, concentrating on the motion of the individual beads. Depending on the bead
position along the chain we found a very rich structure, characterized by
fairly broad successive crossover regimes. Initial effects due to the discreteness of
the model die out very slowly. For chains of length $N \alt 100$ they are
visible for the central bead up to times of the order of the Rouse time, and
for the end beads they are even more pronounced.

Our model assumes the smallest possible tube diameter in the sense that chain segments between two successive entanglement points show no internal
structure. Even in the nonuniversal initial range the results therefore should be relevant for the quantitative interpretation of simulations of
lattice chains in surroundings of high obstacle density. Indeed, as shown in
\cite{Z12} and in our subsequent paper \cite{Z11}, they perform very well in
explaining data of the Evans-Edwards model. A glance to Fig.\ 8 of ref. \cite{Z9} or Fig.\ 6 of ref. \cite{Z19} reveals that also more complicated
dense lattice or continuum systems for the motion of the central bead of the
chain yield results closely resembling our findings. We therefore believe
that even in the microstructure dependent range our results quite generally
show the typical effects of reptational motion.

There remains the question which features could be used to distinguish reptation from motion dominated by disorder or affected by relaxation of the
environment. As we have seen, the unambiguous identification of the power laws
will be most difficult. For values $N = O(100)$, as typically reached in
simulations, only the motion of the central bead shows a small region described by the $t^{1/4}$-law, and to unambigously identify this behavior as
a limiting law its stability with increasing chain length would have to be
checked. Beads not close to the center of the chain feel successive crossovers which essentially wash out all the $t^{1/4}$-structure. However,
as pointed out at the end of section 4 and illustrated in \cite{Z11,Z12},
appropriate fourth moments of ${\bf r}_{j}(t) - {\bf r}_{j}(0)$ show the
$t^{1/4}$ behavior much more clearly, even for very short chains. Such moments therefore seem to be the appropriate objects in a search for the
initial $t^{1/4}$ behavior. At least, a comparison of the second and fourth moments will allow us to identify the range and magnitude of the nonuniversal initial effects.

According to reptation theory the $t^{1/4}$-regime for $T_{2} \ll t \ll T_{3}$ is followed by $t^{1/2}$-behavior. Our results show that for typical
chain lengths this interval is too small to identify such behavior, in particular if we take into account the fairly strong fluctuations in the tube
renewal $(n_{max}(t))$. In practice we for $t > T_{2}$ expect to see a smooth crossover towards free diffusion. To find indications of the $t^{1/2}$-regime it will be preferable to measure the motion of the central
bead relative to the center of mass: $g_{2}(\frac{N}{2},N,t) = \left\langle
(\overline{{\bf r}_{N/2}(t) - {\bf R}_{cm}(t) - {\bf r}_{N/2}(0) + {\bf R}_{cm}(0)})^{2}\right\rangle$. This quantity for $t \ll T_{3}$ shows the
same behavior as $g_{1}(\frac{N}{2},N,t)$, but an eventual increase of the
effective exponent from $x \approx 1/4$ towards $x = 1/2$ cannot be interpreted as crossover towards $x = 1$, since $g_{2}(\ldots)$ for $t >
T_{3}$ crosses over to a constant $g_{2}(\frac{N}{2},N,t) \approx R_{g}^{2}$.

As a further quite pronounced signature of reptation we note the difference
in the behavior of the end bead as compared to the central bead of the chain.
For $t \ll T_{2}$ the ratio $g_{1}(0,N,t)/g_{1}(\frac{N}{2},N,t)$ in the
Rouse model tends to $2$, and we do not expect this to be strongly modified
by disorder, which in leading approximation just renormalizes the over all
time scale \cite{Z7,Z20}. In contrast, we in our mean hopping rate approximation find a much larger asymptotic value $4 \sqrt{2}$ of this ratio.
This not only illustrates the effectiveness of the tube in constraining the
motion, but also it is a consequence of the slow decay of the correlations of
$n_{max}(t)$ in time implied by the basic defect motion. We believe that
this enhancement of $g_{1}(0,N,t)/g_{1}(\frac{N}{2},N,t)$ should be observable fairly easily.

\vspace{2cm}

{\Large Acknowledgement}

This work was supported by the Deutsche Forschungsgemeinschaft, SFB `Unordnung und grosse Fluktuationen'. Furthermore financial support of UE by the Dutch research foundation NWO and by the EU-TMR-network `Patterns, Noise and Chaos' is gratefully acknowledged. 

\newpage
\appendix
\section{Some properties of the coefficients $A_{b}$}
\subsection{$A_{1}(j,t)$}

Eq.\ (3.12) defines
\begin{equation}
A_{1} (j,t) = \frac{pt}{N} + \frac{1}{2N} \sum_{k=1}^{N-1}\:(1 -
\alpha_{k}^{t})\: \frac{\cos^{2}\left(\frac{\pi k}{N} \left(j +
\frac{1}{2}\right)\right)}{\sin^{2} \left(\frac{\pi k}{2N}\right)} \: \: \:.
\end{equation}
Evaluating the first contribution in the sum we find the equivalent form
\begin{eqnarray}
A_{1} (j,t) &=& \frac{pt}{N} + \frac{N}{3} - \frac{1}{2} + \frac{1}{6N} -
\left(1 - \frac{1}{N}\right)\:j + \frac{j^{2}}{N}
\nonumber \\
& & - \frac{1}{2N} \sum_{k=1}^{N-1}\:\alpha_{k}^{t}\:
\frac{\cos^{2}\left(\frac{\pi k}{N} \left(j +
\frac{1}{2}\right)\right)}{\sin^{2} \left(\frac{\pi k}{2N}\right)}\: \: \:.
\end{eqnarray}
Defining $\bar{A}_{1}(\bar{j},\bar{t}) = A_{1}(j,t)/N; \bar{j} = j/N, \bar{t} = p t/N^{2}$ we can take the continuous chain limit $N \rightarrow
\infty, \bar{t},\bar{j}$ fixed, to immediately arrive at the results
\begin{equation}
\bar{A}_{1}(\bar{j},\bar{t}) = \bar{t} + \left(\bar{j} -
\frac{1}{2}\right)^{2} + \frac{1}{12} -
\frac{2}{\pi^{2}}\:\sum^{\infty}_{k=1}\:\frac{\cos^{2}(\pi k
\bar{j})}{k^{2}}\:e^{- \pi^{2} k^{2} \bar{t}}
\end{equation}
\begin{equation}
= \frac{1}{2}
\sqrt{\frac{\bar{t}}{\pi}}\:\int_{0}^{1}\:\frac{ds}{\sqrt{s}}\:\sum^{+ \infty}_{\nu = - \infty}\:\left[e^{- \frac{\nu^{2}}{\rule{0mm}{1ex}\bar{t}s}}
+ e^{- \frac{(\nu - \bar{j})^{2}}{\rule{0mm}{1ex}\bar{t} s}}\right] \: \:
\:.
\end{equation}
(For comparable calculations in more detail, see the appendices of \cite{Z20}.)
Besides yielding the asymptotic limits $\bar{t} \rightarrow 0,\infty$, these
expressions allow for an efficient numerical evaluation. Up to errors less
than $10^{-15}$ we may use the first form truncated at $k=3$ for $\bar{t}
\geq 0.2$, and the second form truncated at $|\nu| = 2$ for $\bar{t} \leq
0.2$.

In the limit $N \rightarrow \infty,t,\bar{j}$ fixed, we go back to the form
A.1 written as
\begin{eqnarray*}
A_{1} (j,t) &=& \frac{pt}{N} + \frac{1}{4N} \sum_{k=1}^{N-1}\: \frac{1 -
\alpha_{k}^{t}}{\sin^{2} \left(\frac{\pi k}{2N}\right)} \left[ 1 + \cos \left(
\pi\:k\left(2 \bar{j} + \frac{1}{N}\right)\right)\right]
\nonumber \\
&\stackrel{N \rightarrow \infty}{=}&
\:\frac{1}{2\pi}\:\int_{0}^{\pi/2}\:dx\:\frac{1 - (1 - 4 p \sin^{2} x)^{t}}{\sin^{2}\:x}
\nonumber \\
& & +\: \frac{1}{4 N} \sum_{k=1}^{N-1}\: \frac{1 - \alpha_{k}^{t}}{\sin^{2} \left(\frac{\pi k}{2N}\right)}\: \cos \left( \pi\:k\left(2 \bar{j} + \frac{1}{N}\right)\right) \Big|_{N \rightarrow \infty}
\end{eqnarray*}
The last contribution for $N \rightarrow \infty$ oszillates to zero. The
first term by partial integration can be written as
\begin{equation}
A_{1}(j,t) = \frac{4}{\pi}\:pt\:\int_{0}^{\pi/2}\:dx\:\cos^{2}\:x \left(1 - 4
p\:\sin^{2}\:x\right)^{t-1}
\end{equation}

\subsection{$A_{2}(s)$}

Eq.\ (3.14) defines
\begin{equation}
A_{2}(s) = \frac{2}{N}
\sum^{N-1}_{k=1}\:\alpha_{k}^{s}\:\sin^{2}\:\left(\frac{\pi k}{N}\right)
\equiv G_{11}^{(0)} (s) \: \: \:.
\end{equation}
It is straightforward to derive the relation
\begin{eqnarray*}
\sum_{0 < s_{1} \leq s_{2} < t}\:A_{2}(s) = \frac{t}{p} \left( 1 - \frac{1}{N}\right) - \frac{1}{2 p^{2} N} \sum^{N-1}_{k=1}\:(1 -
\alpha_{k}^{t})\:\frac{\cos^{2}\left(\frac{\pi k}{2 N}\right)}{\sin^{2} \left(\frac{\pi k}{2 N}\right)} \: \:,
\end{eqnarray*}
leading to
\begin{equation}
p t - p^{2} \sum_{0 < s_{1} \leq s_{2} < t}\:A_{2}(s) = A_{1}(0,t) \: \: \:.
\end{equation}
For $s \gg N^{2}, A_{2}(s)$ vanishes as $\exp \left(- \pi^{2}
\frac{ps}{N^{2}}\right)$. In the limit $N \rightarrow \infty, s$ fixed, $\bar{A}_{2}(s) = A_{2}(s)/N$ reduces to the integral
\begin{equation}
\bar{A}_{2}(s) = \frac{2}{\pi}\:\int_{0}^{\pi}\:dx\:\sin^{2}\:x \left[ 1 - 4
p\:\sin^{2} \left(\frac{x}{2}\right)\right]^{s} \: \: \:,
\end{equation}
which for large $s$ behaves as
\begin{equation}
\bar{A}_{2}(s) \stackrel{s \gg 1}{\simeq}
\:\frac{1}{2 \sqrt{\pi}}\:(p s)^{- 3/2} \: \: \:.
\end{equation}
This illustrates the long range of the correlations of the stochastic process $n(0,s)$.

\subsection{$A_{3}(j,s)$}

Eq.\ (3.16) defines
\begin{equation}
A_{3}(j,s) = \frac{1}{N} + \frac{2}{N} \sum^{N-1}_{k=1}\:\alpha_{k}^{s}\:\cos
\left(\frac{\pi k}{2N}\right) \cos \left(\frac{\pi k}{N} \left(j + \frac{1}{2}\right)\right) \: \: \:.
\end{equation}
We mainly are interested in the sum
\begin{equation}
\tilde{A}_{3}(j,t) = p \left(\frac{t - 1}{N} +
\sum^{t-1}_{s=1}\:A_{3}(j,s)\right) \: \: \:.
\end{equation}
Straightforward summation yields
\begin{eqnarray}
\tilde{A}_{3}(j,t) &=& \frac{p t}{N} + \frac{N}{3} - \frac{1}{2} + \frac{1}{6N} + \frac{j^{2}}{2 N} - \left(1 - \frac{1}{2 N}\right)\:j
\nonumber \\
&-& \frac{1}{2N} \sum^{N-1}_{k=1}\:\alpha_{k}^{t}\:\frac{\cos \left(\frac{\pi
k}{2N}\right)\:\cos \left(\frac{\pi k}{N} \left(j +
\frac{1}{2}\right)\right)}{\sin^{2} \left(\frac{\pi k}{2 N}\right)} \: \: \:,
\end{eqnarray}
where we used the relation
\begin{equation}
\sum^{N-1}_{k=1}\:\cos\:\left(\frac{\pi k}{2N}\right)\:\cos\:\left(\frac{\pi
k}{N} \left(j + \frac{1}{2}\right) \right) = - \frac{1}{2}
\end{equation}
to simplify the result. Comparing to Eq.\ (A.2) we find
\begin{eqnarray*}
\tilde{A}_{3}(0,t) = A_{1}(0,t) \: \: \:,
\end{eqnarray*}
as it should. For $N \rightarrow \infty, \bar{t},\bar{j}$ fixed,
$\hat{A}_{3}(\bar{j},\bar{t}) = \frac{1}{N} \tilde{A}_{3}(j,t)$ takes the
form
\begin{equation}
\hat{A}_{3}(\bar{j},\bar{t}) = \bar{t} + \frac{1}{3} + \frac{\bar{j}^{2}}{2}
- \bar{j} - \frac{2}{\pi^{2}}\:\sum^{\infty}_{k=1}\:\frac{\cos (\pi k \bar{j})}{k^{2}}\:e^{- \pi^{2} k^{2} \bar{t}} \: \: \:.
\end{equation}
Thus
\begin{equation}
\hat{A}_{3}(\bar{j},\bar{t}) \stackrel{\bar t\gg 1}{\simeq}\bar{A}_{1} (0,\bar{t}) + const = \bar{A}_{1}(\bar{j},\bar{t}) + const \: \: \:.
\end{equation}
For $\bar{j} > 0, \bar{t} \rightarrow 0, \hat{A}_{3}$ vanishes exponentially.
\begin{equation}
\hat{A}_{3}(\bar{j},\bar{t})\stackrel{\bar t\ll 1}{\simeq} 
\frac{4}{\sqrt{\pi}} \frac{\bar{t}^{3/2}}{\bar{j}^{2}}\:e^{- \bar{j}^{2}/4
\bar{t}} \: \: \:.
\end{equation}

\section{Spitzer's identity and end bead motion}

For an uncorrelated discrete stochastic process $n_{s}$, Spitzer \cite{Z17}
has derived an expression for the generating function of the simultaneous
distribution of $n_{t}$ and
\begin{equation}
n_{max}(t) = \max_{s \epsilon[0,t]}
\:n_{s} \: \: \:.
\end{equation}
(It is understood that $n_{0} = 0$, by definition.) For a process symmetric
in its hopping probabilities, the result reads
\begin{eqnarray}
\sum^{\infty}_{t=0}\: &\lambda^{t}&\: \overline{z^{n_{max}(t)}
w^{n_{max}(t) - n_{t}}}
\nonumber \\
&=& (1-\lambda)^{-1}\:\exp \Big\{
\sum^{\infty}_{s=1}\:\frac{\lambda^{s}}{s}\: \overline{\left(z^{n_{s}} +
w^{n_{s}} - 2\right) \Theta (n_{s} - 1)}\Big\} \: \: \:.
\end{eqnarray}
Taking the derivative with respect to $z$ and putting $z = 1 = w$, we immediately find
\begin{equation}
\sum^{\infty}_{t=0}\: \lambda^{t}\: \overline{n_{max}(t)} = (1 -
\lambda)^{-1} \sum^{\infty}_{s=1}\:\frac{\lambda^{s}}{s}\:\overline{n_{s}
\Theta (n_{s} - 1)} \: \: \:.
\end{equation}
Inverting the discrete Laplace transformation we find the result (5.1):
\begin{eqnarray*}
\overline{n_{max}(t)} = \sum^{t}_{s=1}\:\frac{\overline{|n_{s}|}}{2 s}
\end{eqnarray*}

In the same way we can construct an approximation for the fourth moment
\begin{equation}
\hat{g}_{1} (0,N,t) = \left\langle \overline{\sum^{3}_{\alpha =
1}\:{(r_{0,\alpha}(t) - r_{0,\alpha}(0))}^{4}}\right\rangle^{1/2} \: \: \:.
\end{equation}
Following the argument of Sect.~II.A we find the expression
\begin{eqnarray}
\hat{g}_{1}^{2} (0,N,t) &=& \overline{\left(2 \ell_{s} n_{max}(t) - \ell_{s} n(0,t)\right)^{2}}
\nonumber \\
&=& \ell_{s}^{2} \left(\overline{4 n^{2}_{max}(t) - 4 n_{max}(t) n(0,t) +
n^{2} (0,t)}\right) \: \: \:.
\end{eqnarray}
Again replacing $n(0,s)$ by an uncorrelated process $n_{s}$, we use Eq.\ (B
2) to evaluate this expression. We find
\begin{equation}
\overline{n^{2}_{max}(t)} = \sum^{t}_{s=1}\:\frac{\overline{n_{s}^{2}}}{2 s} +
\sum^{t-1}_{s,s'=1}\:\frac{\overline{|n_{s}|}\;\overline{|n_{s'}|}}{4 s s'}\:\Theta (t - s - s')
\end{equation}
\begin{equation}
\overline{n_{max}(t) n_{t}} = \sum^{t}_{s=1}\:\frac{\overline{n_{s}^{2}}}{2 s}
\end{equation}
This results in the approximation
\begin{equation}
\hat{g}_{1} (0,N,t) = \ell_{s} \left[
\sum^{t-1}_{s,s'=1}\:\frac{\overline{|n(0,s)|}\;\overline{|n(0,s')|}}{s s'}\:\Theta (t - s - s') + \overline{n^{2} (0,t)}\right]^{1/2}
\end{equation}
Of particular interest is the universal region where all discreteness corrections have died out. Eqs.\ (3.24), (3.26) yield
\begin{eqnarray*}
n^{2}(j,t) &=& 2 \rho_{0} A_{1}(j,t)
\nonumber \\
|n(j,s)| &=& \frac{2}{\sqrt{\pi}}\: \left(\rho_{0} A_{1} (j,s)\right)^{1/2}
\: \: \:.
\end{eqnarray*}
For times large compared to $T_{2}$ we can use the approximation
(cf.\ Eq.\ 5.23)
\begin{eqnarray*}
A_{1} (0,s) = N \bar{s} \: \: \:,
\end{eqnarray*}
and we can evaluate the summations in Eq.\ (B.8) as integrals. This yields
\begin{equation}
\frac{\hat{g}_{1}(0,N,t)}{g_{1}(0,N,t)} = \sqrt{\frac{3}{4}\:\pi}\: \approx
1.085, \: \: \: \: \: T_{2} \ll t \ll T_{3} \: \: \:.
\end{equation}
For $T_{0} \ll t \ll T_{2}$ we find (cf.\ Eq.\ 5.23)
\begin{eqnarray*}
A_{1}(0,s) = 2 N \sqrt{\frac{s}{\pi}} \: \: \:,
\end{eqnarray*}
which in the same way leads to
\begin{equation}
\frac{\hat{g}_{1}(0,N,t)}{g_{1}(0,N,t)} = \frac{1}{4} \left(\frac{\pi}{2} +
\frac{2}{\pi^{1/2}}\:\Gamma^{2} \left(\frac{1}{4}\right)\right)^{1/4} \approx
1.0125, \: \: \: \: \: T_{0} \ll t \ll T_{2}
\end{equation}
This should be compared to the corresponding result for the central segment
\begin{equation}
\frac{\hat{g}_{1}(\frac{N}{2},N,t)}{g_{1}(\frac{N}{2},N,t)} =
\sqrt{\frac{\pi}{2}}\:\approx 1.253, \: \: \: T_{0} \ll t \ll T_{3} \: \: \:.
\end{equation}
For the end segment the ratio $\hat{g}_{1}/g_{1}$ in all the universal time
region is much closer to $1$.

\section{Approximate form of}
\vspace{-.5cm}
\begin{center}
$\textstyle \hat{{\cal P}}_{max,j}(n_{m},n;t)$
\end{center}

To construct an approximation for the simultaneous distribution of $n_{max}(t)$ and $n(j,t)$ we start from the factorization (5.40) to write
\begin{equation}
\hat{{\cal P}}_{max,j}(n_{m},n;t) = \sum^{+ \infty}_{n_{t} = - \infty}
\hat{{\cal P}}_{max,0}(n_{m}|n_{t};t) \hat{{\cal P}}_{2}(n_{t},n;0,j,t)
\end{equation}
A rigorous expression for $\hat{{\cal P}}_{2}(n_{t},n;0,j,t)$ is given in
Eq.\ (3.30). For the conditional probability $\hat{{\cal
P}}_{max,0}(n_{m}|n_{t};t)$ to find $n_{max}(t) = n_{m}$ for given $n(0,t) =
n_{t}$ we take the result of a simple random walk, making steps $\pm 1$ with
probability $p_{0}$ and steps $0$ with probability $1-2 p_{0}$.

Let the walk start at $n_{0}=0$, reach $n_{max}(t) = n_{m}$ at time $s$ for the
first time, and end at $n_{t} \leq n_{m}$. The contributions of the parts
$[0,s-1]$ or $[s,t]$ of this process can be written in terms of powers
of appropriately truncated matrices of type $\hat{W}$ (Eq.\ 3.7), and the
evaluation is straightforward. We find
\begin{eqnarray}
\hat{{\cal P}}_{1}(n_{t};t) \hat{{\cal P}}_{max,0}(n_{m}|n_{t};t) &=& \Theta
\left(n_{m} - \frac{1}{2} (n_{t} + |n_{t}|)\right) (1 - \delta_{n_{m}0})
p_{0}
\nonumber \\
& & \cdot \frac{4}{\pi^{2}}\:\int_{0}^{\infty}\:dx_{1} dx_{2}\:\sin (x_{1} (1 +
n_{m} + n_{t})) \sin (x_{1}) \sin (x_{2}) \sin (x_{2} n_{m})
\nonumber \\
& & \cdot \alpha_{0} (x_{1}) \frac{\alpha_{0}^{t-1}(x_{1}) -  \alpha_{0}^{t-1}
(x_{2})}{\alpha_{0}(x_{1}) - \alpha_{0}(x_{2})}
\nonumber \\
& & + \Theta (n_{m} - 1) \delta_{n_{m}n_{t}}\:p_{0}
\frac{2}{\pi}\:\int_{0}^{\infty}\:dx\:\sin(x) \sin (x n_{m})
\alpha_{0}^{t-1}(x)
\nonumber \\
& & + \delta_{n_{m}0}\:\Theta (- n_{t})
\frac{2}{\pi}\:\int_{0}^{\infty}\:dx\:\sin (x(1-n_{t})) \sin (x)
\alpha_{0}^{t}(x) \: \: \:,
\end{eqnarray}
where
\begin{equation}
\alpha_{0}(x) = 1 - 4 p_{0}\:\sin^{2} \left(\frac{x}{2}\right) \: \: \:.
\end{equation}
We now may adjust $p_{0}$ so as to reach the desired value of
$\overline{n_{max}(t)}$:
\begin{equation}
\sum^{+ \infty}_{n_{t} = - \infty}\:\sum^{\infty}_{n_{m} = 0}\:n_{m} \hat{{\cal P}}_{1}(n_{t};t) \hat{{\cal P}}_{max,0}(n_{m}|n_{t};t) = \overline{n_{max}(t)}
\end{equation}

This construction of $\hat{{\cal P}}_{max,j}(n_{m},n;t)$ guarantees that
\begin{itemize}
\item[i)] the relations $0 \leq n_{m}$ and $n_{t} \leq n_{m}$ are obeyed, 
\item[ii)] $\overline{n^{2}(j,t)}$ and $\overline{n_{max}(t)}$ take their
proper values, and
\item[iii)] $g_{1}(j,N,t)$ (Eq.\ 2.13) evaluated with this distribution for
$j > 0, t < T_{R}(j)$ reduces to the results of Section IV, and for $j=0$
yields $g_{1}(0,N,t)$ as evaluated in Sect.\ V.B..
\end{itemize}

Being interested mainly in beads $j$ not extremely close to a chain end $(j
\agt 20$, say), we note that $p T_{R}(j) \gg 1$, so that we may simplify the
result by taking the limit of large time: $\alpha_{0}^{t}(x) \rightarrow \exp
(- p_{0} t x^{2})$. Then the integrals in Eq.\ (C.2) can be evaluated further
to yield
\begin{eqnarray}
\hat{{\cal P}}_{max,0}(n_{m}|n_{t};t) &=& \frac{\Theta\left(n_{m} - \frac{1}{2} (n_{t} + |n_{t}|)\right)}{\overline{n_{max}(t)}} 2
\sqrt{\frac{2}{\pi}} \left( 2 \sqrt{\frac{2}{\pi}}\:z_{m} -
z_{t}\right)\nonumber\\&&\exp \left[ - \frac{4}{\pi}\:z_{m}^{2} + 2 \sqrt{\frac{2}{\pi}}\:z_{m} z_{t}\right]
 + O \left(\frac{1}{\sqrt{p_{0} t}}\right)
\end{eqnarray}
where we use the notation
\begin{eqnarray}
z_{m} &=& \frac{n_{m}}{\overline{n_{max}(t)}} = \sqrt{\frac{\pi}{p_{0} t}}
\frac{n_{m}}{2}
\nonumber \\
z_{t} &=& \frac{n_{t}}{(\overline{n^{2}(0,t)})^{1/2}} = \frac{n_{t}}{\sqrt{2
p_{0} t}} \: \: \:.
\end{eqnarray}
The neglected terms of relative order $\sqrt{1/p_{0} t}$ are due to boundary
effects (terms $\delta_{n_{m}0}, \delta_{n_{m} n_{t}}$ in Eq.\ (C.2)).

We now evaluate $\hat{{\cal P}}_{2}(n_{t},n;0,j,t)$ in the corresponding
approximation, which implies expanding the argument of the exponential in
Eq.\ (3.30) to second order in $\varphi, \varphi_{t}$. The resulting Gaussian integral yields
\begin{eqnarray}
\hat{{\cal P}}_{2}(n_{t},n;0,j,t) &=& \frac{1}{2\pi}
\left(\overline{n^{2}(0,t)} \overline{n^{2}(j,t)}\right)^{-1/2}\: (1 - a^{2})^{-1/2}
\nonumber \\
&\cdot& \exp \left[ - \frac{1}{2 - 2 a^{2}} \left(z_{t}^{2} + z^{2} - 2 a\:
z\: z_{t}\right)\right] \: \: \:,
\end{eqnarray}
where
\begin{equation}
z = \frac{n}{(\overline{n^{2}(j,t)})^{1/2}}\: \: \:,
\end{equation}
\begin{equation}
a = \frac{\tilde{A}_{3}(j,t)}{(A_{1}(0,t) A_{1}(j,t))^{1/2}} \: \: \:.
\end{equation}
Inserting Eqs.\ (C.5), (C.9) into Eq.\ (C.1) and carrying through the summation over $n_{t}$ as integral over $z_{t}$ we find the result given in
Eq.\ (5.46).

\newpage
{\Large Figure captions}

SORRY: Figures 3 and 5 have an unconventional format and are
not included in this file. For copies of these figures please
contact ebert@lorentz.leidenuniv.nl.

Fig.\ 1 \\
The Evans-Edwards model in 2 dimensions:
a piece of a chain on a square lattice, showing two hairpins. The crosses
indicate the obstacles. The hairpins are the mobile units corresponding to the `defects'.

Fig.\ 2 \\
Reptational motion. The full line gives the initial chain configuration. The
broken lines are the new parts found at time $t$. Arrows point to $j_{<}$ or
$j_{>}$, respectively. The motion of a bead in the original tube is indicated,
by its initial (heavy dot) and final (shaded circle) positions.

Fig.\ 3 \\
$\log_{10} (\bar{g}_{1}(\bar{j},\bar{t})/\bar{g}_{ass}(\bar{t}))$
as function of $\log_{10}\:\bar{t}$ for $\bar{j} = 1/100,\: 1/10,\: 1/2$.
$\bar{g}_{ass}(\bar{t}) = 2 \ell_{s}\: \sqrt{\frac{\rho_{0}}{\pi}}\:\left(\frac{\bar{t}}{\pi}\right)^{1/4}$ represents the first $t^{1/4}$-regime.
Short dashes: height of the second $(\sqrt{2}$-) $t^{1/4}$-regime. Long dashes: $t^{1/2}$-behavior.

Fig.\ 4\\
$g_{1}(N/2,N,t)/g_{ass}(t)$ as function of $\log_{10}(pt)$, where $g_{ass}(t) = \frac{2}{\sqrt{\pi}}\:\left(\tilde{t}/\pi\right)^{1/4}$, cf. Eq.~(4.13). 
Full line: $N \rightarrow \infty$, long dashes: $N = 40,\: 100,\: 1000$, short dashes: The $t^{1/4}$-asymptote, here normalized to height 1 by extracting $g_{ass}(t)$. 

Fig.\ 5 \\
Ratio $\bar{g}_{1}(0,\bar{t})/\bar{g}_{1}(\frac{1}{2},\bar{t})$ as function
of $\log_{10} \bar{t}$.

Fig.\ 6 \\
$g_{1}(0,N,t)/g_{ass}(0,t)$ asymptote as function of $\log_{10}(pt)$. $g_{ass}(0,t)$ is given by Eq.~(5.28). Full line: $N \rightarrow \infty$. Long dashes: $N = 40,\:100,\:1000$. The short dashes give the $t^{1/4}$-asymptote  here
found at $1$ by normalization.

Fig.\ 7 \\
Schematic plot of $g_{1}(j,N,t)/g_{1}(\frac{N}{2},N,t)$ for $1 \ll j \ll N$,
showing the sequence of crossover behavior discussed in the text. Division
by $g_{1}(\frac{N}{2},N,t)$ serves to transform power law regimes to plateaus
indicated by the thin lines.

Fig.\ 8 \\
Simultaneous distribution of $n_{max}(t),n(j,t)$ for values
$n_{max}(t)/\overline{n_{max}(t)} = z_{m} = 1,2,3$ as function of $z = n(j,t)/(\overline{n^{2}(j,t)})^{1/2}$. Full lines: $a(j,t) = 0.9$; broken
lines: $a(j,t) = 0.99$.

Fig.\ 9 \\
Results for $g_{1}(j,N=10^{3},t)$ normalized to the initial $t^{1/4}$-behavior
$g_{ass}(t) = \frac{2}{\sqrt{\pi}}\:\left(\tilde{t}/\pi\right)^{1/4}$, cf. Eq.~(4.13),
for a) $j=20$; b) $j = 100$. Fat full lines give the full result. Thin
lines: $g_{i}(\ldots)$. Dot-dashed: $g_{r}(\ldots)$. Dotted: $g(0,10^{3},t)$.
The arrows indicate $T_{R}(j)$.

Fig.\ 10 \\
Reptation time as function of chain length. Fat line: full model. Thin line:
continuous chain limit. Long dashes: $T_{3} \sim N^{3}$. Short dashes: $T_{3} \sim N^{3.4}$.
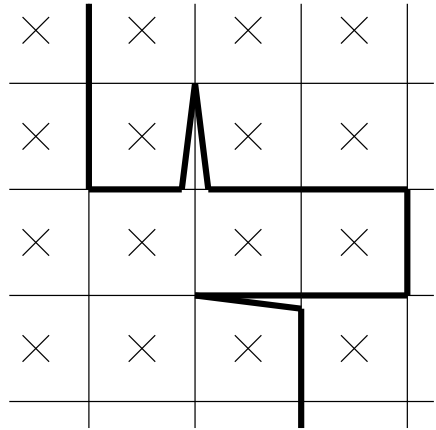
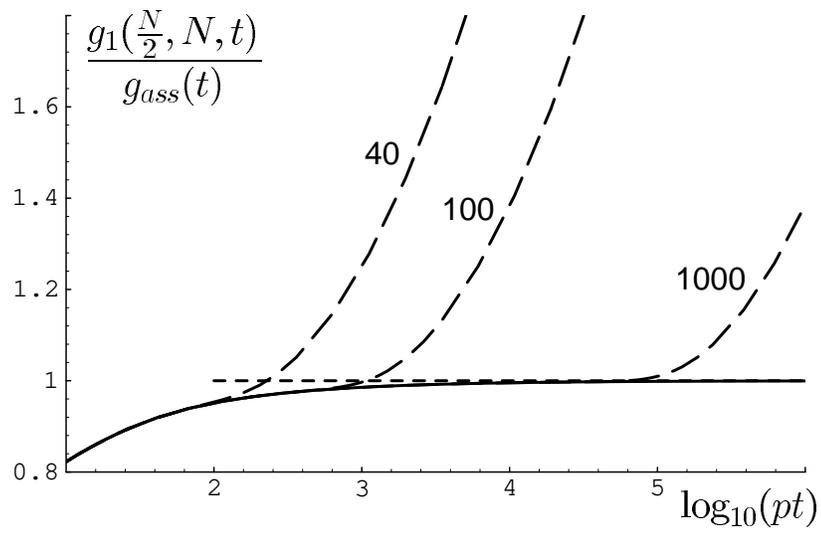
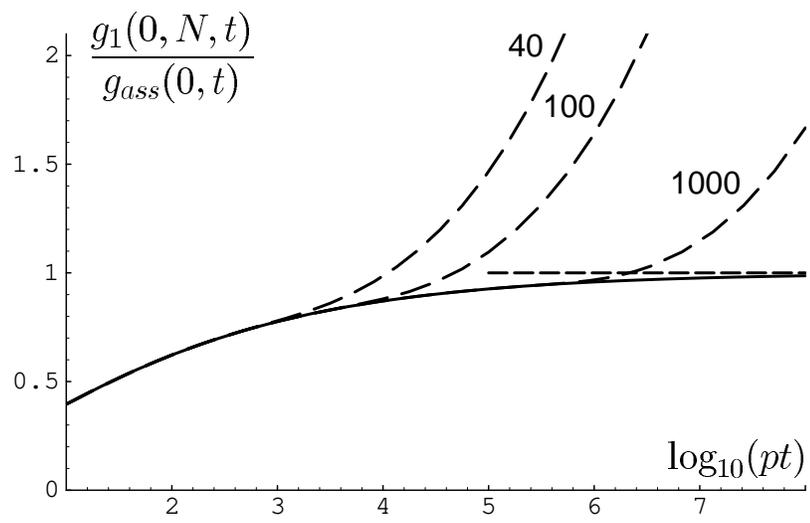
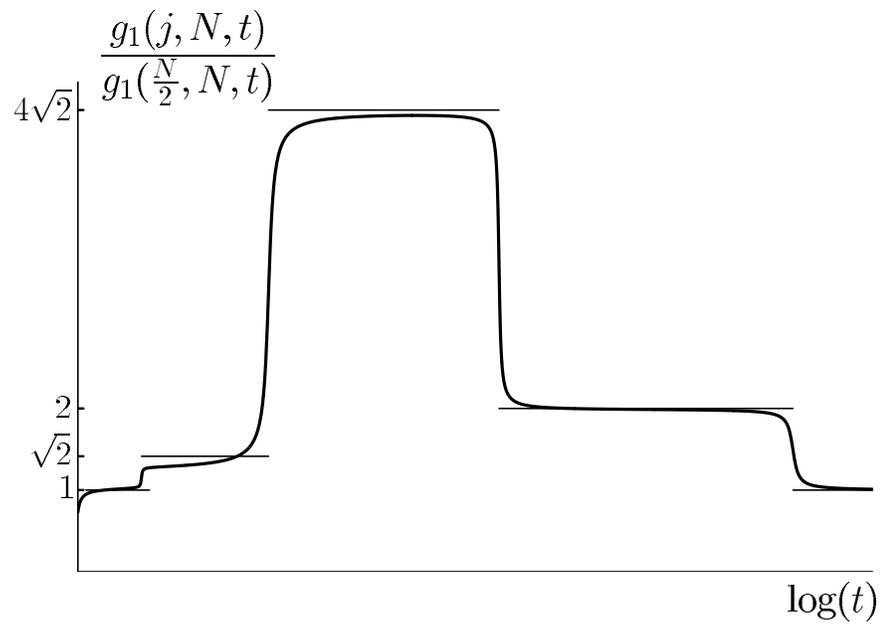
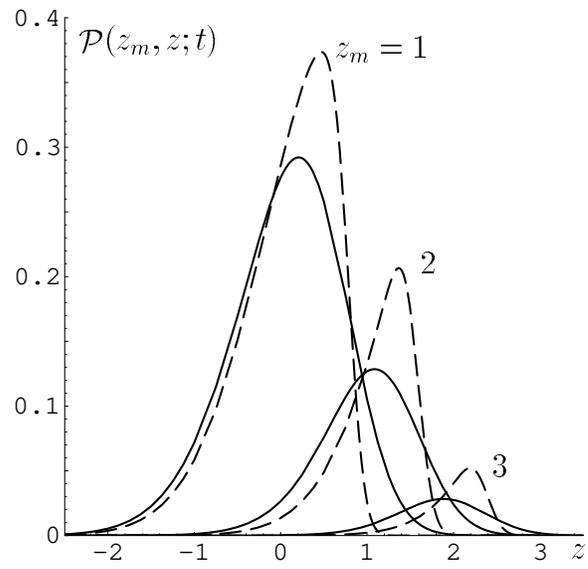
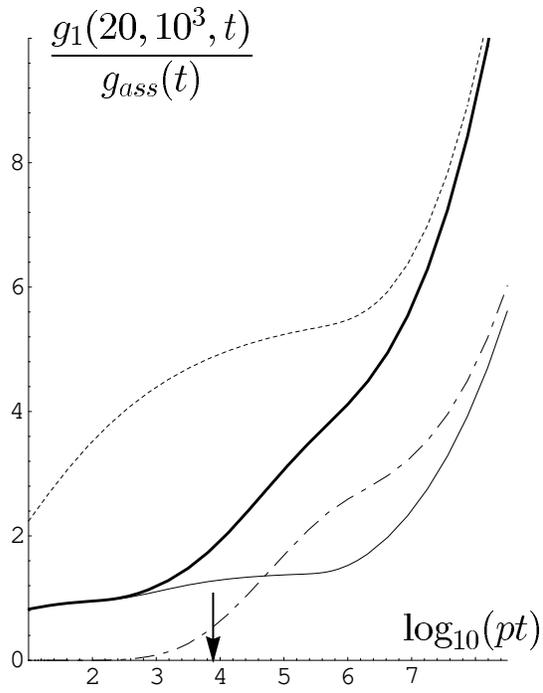
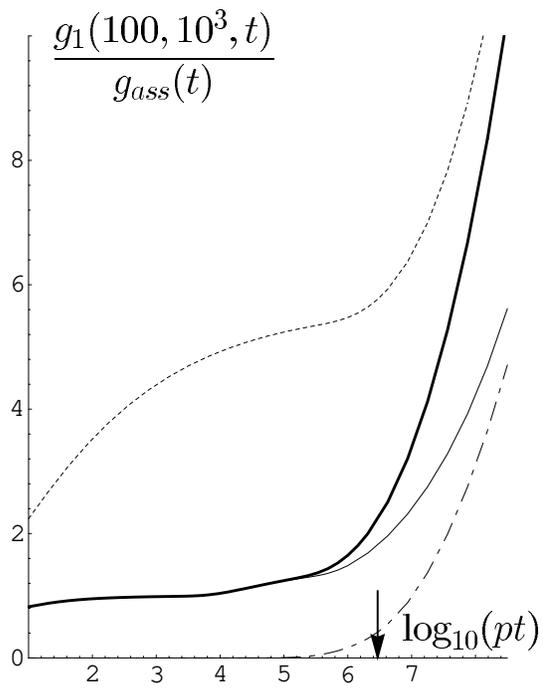
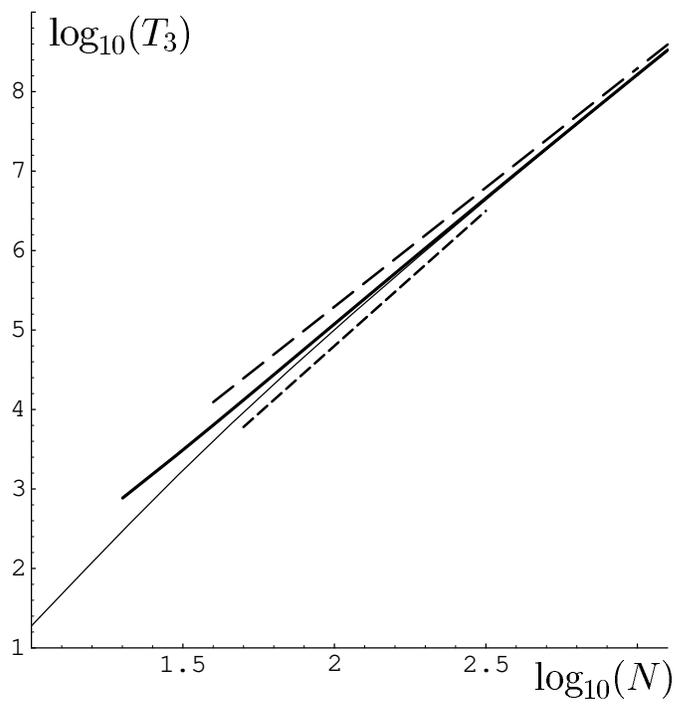
\end{document}